\documentclass[a4paper,12pt]{article}

\usepackage{amsmath}
\usepackage{amsfonts}
\usepackage{amssymb}
\usepackage{graphicx}
\numberwithin{equation}{section}

\newcommand{\de}{\partial}

\newcommand{\del}{\delta}

\renewcommand{\th}{\theta}

\newcommand{\Lam}{\Lambda}

\newcommand{\ud}{\textrm{d}}
\newcommand{\nab}{\nabla}
\newcommand{\Mb}{\bar{M}}
\newcommand{\Mpl}{M_{\textrm{Pl}}}
\newcommand{\Mp}{M_{\textrm{Pl}}}

\begin{document}

\def\thefootnote{\fnsymbol{footnote}}

\begin{center}
\Large{\textbf{The Effective Theory of Quintessence: \\
the $w<-1$ Side Unveiled}}
\\[0.5cm]
\large{Paolo Creminelli$^{\rm a}$, Guido D'Amico$^{\rm b,c}$,\\[.1cm]
Jorge Nore\~na$^{\rm b}$, and Filippo Vernizzi$^{\rm d,a}$}
\\[0.5cm]

\small{
\textit{$^{\rm a}$ Abdus Salam International Center for Theoretical Physics\\ Strada Costiera 11, 34014, Trieste, Italy}}

\vspace{.2cm}

\small{
\textit{$^{\rm b}$ SISSA/ISAS, via Beirut 2-4, 34014, Trieste, Italy}}

\vspace{.2cm}

\small{
\textit{$^{\rm c}$ INFN - Sezione di Trieste, 34014 Trieste, Italy}}

\vspace{.2cm}

\small{
\textit{$^{\rm d}$ CEA, IPhT, 91191 Gif-sur-Yvette c\'edex, France\\ CNRS, URA-2306, 91191 Gif-sur-Yvette c\'edex, France}}

\end{center}

\vspace{.8cm}

\hrule \vspace{0.3cm}
\noindent \small{\textbf{Abstract}} \\[0.3cm]
\noindent
We study generic single-field dark energy models, by a
parametrization of the most general theory of their perturbations around
a given background, including higher derivative terms. In appropriate
limits this approach reproduces standard quintessence, $k$-essence and
ghost condensation.  We find no general pathology associated to an
equation of state $w_Q < -1$ or in crossing the phantom divide $w_Q =
-1$. Stability requires that the $w_Q < -1$ side of dark energy
behaves, on cosmological scales, as a $k$-essence fluid with a
virtually zero speed of sound.  This implies that one should set the
speed of sound to zero when comparing with data models with $w_Q<-1$
or crossing the phantom divide.  We summarize the theoretical and
stability constraints on the {\em quintessential plane} $(1+w_Q) $
vs.~speed of sound squared.

\vspace{0.5cm} \hrule
\def\thefootnote{\arabic{footnote}}
\setcounter{footnote}{0}






\section{Introduction}
The origin of the present acceleration of the Universe is likely to be
the most important theoretical problem in physics today. Given the general
reluctance in accepting as explanation an incredibly small
cosmological constant and the absence of compelling alternatives, it
seems that one should keep an open-minded approach, concentrating on very
general theoretical constraints and on observables more than on
specific models. 

In this paper we study in generality, and
focusing on perturbations, dark energy scenarios where the dark sector
is described by a single scalar degree of freedom, without direct
coupling to matter (in the Einstein frame).  We will often call this
general model {\em quintessence}, although in the literature this name
is usually reserved to a scalar field with a canonical kinetic term.

Following \cite{StartingUniverse,EFTinflation}, we
will rewrite the scalar field Lagrangian in order to make explicit
what is the most general theory of quintessence perturbations around a
given background solution characterized by its pressure $p_Q$ and
energy density $\rho_Q$.  The freedom that we have after the
unperturbed history is fixed is made clear in this way. This
separation is particularly important given that a host of new
experiments is going to test dark energy clustering properties
\cite{Albrecht:2006um}.
In our formalism, the general theoretical
constraints on single field models are also made clear.

In particular, we will study
whether a single field model that is safe from ghost and gradient instabilities
can have an equation of state $w_Q < -1$,
where $w_Q = p_Q/\rho_Q$. In this regime, the stability of the model
can be guaranteed by the presence of higher derivative operators, a conclusion already
    reached in \cite{StartingUniverse}, where single field models were studied focusing on the
    constraints enforced by stability. Here, after reviewing and extending
    the results of \cite{StartingUniverse}, we will concentrate on the behaviour of
    cosmological perturbations, which are relevant for observations. On
    cosmological scales we find that these higher derivative terms are
    irrelevant for the phenomenology, so that a model with
$w_Q < -1$
simply behaves as a $k$-essence fluid with virtually zero speed of
sound. Higher derivative terms are relevant for cosmology only when
the equation of state gets very close (and experimentally
indistinguishable from) a cosmological constant. In this limit our
general Lagrangian reduces to the Ghost Condensate theory
\cite{GhostCond} smoothly connecting quintessence to this theory of
modification of gravity. Notice that, as detailed in Appendix \ref{app:scaling},
we are interested in a regime where higher derivative terms do not
introduce additional degrees of freedom (contrary to what happens for
example in \cite{Li:2005fm}). We find it convenient to summarize our
results in the plane $(1+w_Q) \Omega_Q$ vs.~$c_s^2$, where $\Omega_Q$
is the quintessence contribution to the critical density.
We dub this plane of parameters the
{\em quintessential plane}.
 
We also study the issue of whether it is possible to cross the
so-called phantom divide $w_Q = -1$ \cite{Feng:2004ad,HuPhantom}. We find that the
speed of sound vanishes exactly at the divide
\cite{Vikman:2004dc,CaldwellDoran} and since quintessence may remain
stable for $w_Q<-1$ there is no general pathology associated with the
crossing. We show this explicitly with an example. The phantom divide
can be crossed with a single scalar degree of freedom,
without introducing ghost-like fields.

The paper is organized as follows. In section~\ref{sec:effective} we
study the most general theory of single field quintessence, taking
into account higher derivative operators and focusing on the stability
constraints following \cite{StartingUniverse}. In section~\ref{sec:plane} we study the phenomenology in
various limits, considering also the gravitational effect of dark
matter on quintessence. In section~\ref{sec:divide} we consider the
issue of crossing the phantom divide $w_Q =-1$ and we show explicit
examples of the crossing without pathologies. In
section~\ref{sec:addhigh} we concentrate on another kind of higher
derivative operators \cite{StartingUniverse,EFTinflation}, different from the ones
studied for the Ghost Condensate.  Although the phenomenology on
cosmological scales does not change, the modification of gravity at
short distances is quite different.  Conclusions are drawn in
section~\ref{sec:conclusions}.

Several issues concerning our effective theory approach are left to
the appendix. Appendix~\ref{app:scaling} is devoted to reviewing how
higher derivative operators must be treated in the effective field
theory approach. In appendix~\ref{app:SE} we write down the
stress-energy tensor for the action discussed in the main text.  In
appendix~\ref{app:unitary} we derive the most general action for
quintessence perturbations following the approach of
\cite{StartingUniverse,EFTinflation}. 
Finally, in
appendix~\ref{app:modgrav} we discuss the modification of gravity 
induced by the kind of higher derivative operators that were not 
studied in \cite{GhostCond}.



\section{Effective theory of quintessence}
\label{sec:effective}

Our aim is to study the most general theory of quintessence
perturbations. We will do it step by step, first by considering a
model with an action containing at most a single derivative acting on
the field.  This is known as $k$-essence
\cite{ArmendarizPicon:1999rj,ArmendarizPicon:2000dh} and it will be
possible to write the action for the perturbations in such a way as to
make explicit the dependence on the background energy density and
pressure $\rho_Q$ and $p_Q$. Then we will add higher derivative
operators to the $k$-essence action in such a way as to leave the
background invariant. In this section we will consider the kind of
operators introduced in the context of ghost condensation.  Other
higher derivative operators will be discussed later in
section~\ref{sec:addhigh}.

An alternative derivation of the most general action for quintessence
perturbations is given in appendix~\ref{app:unitary} following the
approach of refs.~\cite{StartingUniverse,EFTinflation}, that consists
in writing down all the terms preserving the symmetries of the system
in a `unitary' gauge, where the quintessence perturbation is set to zero
and appears as a scalar metric degree of freedom.  This approach is
elegant and straightforward but less pedagogical than the one adopted
here. Both approaches lead to the same physical results.


\subsection{The limit of {\em k}-essence}
\label{actionsection}

Let us start with a $k$-essence action
\begin{equation}
S = \!\int \!\ud^4 \, x \sqrt{-g} \;P(\phi,X) \;,
\qquad X = - g^{\mu \nu} \de_\mu \phi \de_\nu \phi \;.
 \label{kaction}
\end{equation}
We assume a flat Friedmann-Lema\^{i}tre-Robertson-Walker (FLRW)
Universe with metric $\ud s^2 = - \ud t^2 + a^2(t) \ud \vec x^2$.  Initially,
we will treat this as a fixed background and neglect the perturbations
of the metric.

To describe perturbations around a given background solution
$\phi_0(t)$, it is useful to write the scalar field as
\begin{equation}
 \phi(t,\vec x) = \phi_0(t + \pi(t,\vec x)) \;,
\end{equation}
and expand the action
(\ref{kaction}) in terms of $\pi$. In the following, we are going to
assume that the function $\phi_0(t)$ is strictly monotonic, $\dot
\phi_0(t)\neq 0$, to avoid singularities in the relation between
$\phi$ and $\pi$.

Using the expansions
\begin{eqnarray}
 \phi(t, \vec x) & = & \phi_0 + \dot{\phi_0} \pi + \frac{1}{2}
 \ddot{\phi}_0 \pi^2 + \ldots \;,\\
 \label{X}
 X(t, \vec x) & = & X_0 + \dot{X_0} \pi + \frac{1}{2} \ddot{X_0} \pi^2 + 2 X_0 \dot{\pi}
  + 2 \dot{X_0} \pi \dot{\pi} + X_0 \left( \dot{\pi}^2  - \frac{(\nab \pi)^2}{a^2}
     \right) + \ldots \;,
\end{eqnarray}
where $X_0 = \dot{\phi_0}^{\!2}$, we have, up to second order in $\pi$,
\begin{equation}
  S  = \!\int \!\ud^4 x \,a^3
  \left[ P_0 + \dot{P_0} \pi + \frac{1}{2} \ddot{P_0} \pi^2
    + 2 P_X X_0 \dot{\pi} + 2 \left(P_X X_0 \right)\dot{} \, \pi \dot{\pi}
    + P_X X_0 \left(  \dot{\pi}^2 -\frac{(\nab \pi)^2}{a^2}   \right)
    + 2 P_{XX} X_0^2 \dot{\pi}^2 \right] \; ,
\label{kessence2nd}
\end{equation}
where $P_X = \partial P /\partial X |_0$ and $P_{XX} = \partial^2 P
/\partial X^2 |_0$.  The term $P_0$ does not affect perturbations as
it is independent of $\pi$, while one can verify that the linear terms
cancel using the background equation of motion.  Indeed, by
integrating by parts the term $\pi \dot \pi$ and making use of
the background equation of motion, after some manipulations we are left with
\begin{equation}
  \label{kessence}
  S = \!\int \!\ud^4 x \, a^3 \Bigg[ \left( P_X X_0 + 2 P_{XX} X_0^2 \right) \dot{\pi}^2
  - P_X X_0 \frac{(\nab \pi)^2}{a^2}
  + 3 \dot{H} P_X X_0 \, \pi^2  \Bigg] \, .
\end{equation}

We can now rewrite the coefficients of this expansion in terms of the
stress-energy tensor of the background solution.
From the definition of the stress-energy tensor,
\begin{equation}
\label{defTmunu}
  T_{\mu \nu} = - \frac{2}{\sqrt{-g}} \frac{\del S}{\del g^{\mu \nu}} \;,
\end{equation}
one obtains the background energy density and pressure,
\begin{equation}
\rho_Q  = 2X_0 P_X - P_0 \;, \qquad p_Q = P_0 \;.
\label{kstressenergy}
\end{equation}
Using these expressions the action above can be cast in the form
\begin{equation}
  \label{kessence2}
  S = \!\int \!\ud^4 x \, a^3
\Bigg[ \frac{1}{2} \left( \rho_Q+p_Q +4 M^4  \right) \dot{\pi}^2
  - \frac{1}{2}(\rho_Q+p_Q) \frac{(\nab \pi)^2}{a^2}
  + \frac{3}{2} \dot{H} (\rho_Q+p_Q)  \pi^2  \Bigg] \; .
\end{equation}
Here we have defined $M^4 \equiv P_{XX}X_0^2$, where $M$ has the
dimension of a mass.

At this stage we can straightforwardly introduce the coupling with
metric perturbations. This coupling is particularly simple in
synchronous gauge where the metric takes the form
\begin{equation}
 \ud s^2 = - \ud t^2 + a^2(t) (\del_{ij} + h_{ij}) \ud x^i \ud x^j \, .
\label{synchronous}
\end{equation}
Indeed, at quadratic order in the action the coupling with gravity only
comes through the perturbed $\sqrt{-g}$ in the action. Replacing $a^3$
with $a^3(1+h/2)$ in eq.~(\ref{kessence2nd}) we have
\begin{equation}
 S  = \!\int \!\ud^4 x \,a^3 \left(1 +\frac{h}{2}\right)
  \Big[  \dot{P_0} \pi + 2 P_X X_0 \dot{\pi} + \ldots \Big] \;.
\end{equation}
Integrating by parts and using again the background equation of motion
one gets the full action for $\pi$,
\begin{equation}
  \label{kessencefull}
  S = \!\int \!\ud^4 x \, a^3
\Bigg[ \frac{1}{2} \left( \rho_Q+p_Q +4 M^4  \right) \dot{\pi}^2
  - \frac{1}{2}(\rho_Q+p_Q) \frac{(\nab \pi)^2}{a^2}
  + \frac{3}{2} \dot{H} (\rho_Q+p_Q)  \pi^2  -
\frac{1}{2}(\rho_Q + p_Q)\dot{h}\pi\Bigg] \; .
\end{equation}

The quadratic Lagrangian is thus specified by the functions $(\rho_Q +
p_Q)(t)$ and $M^4(t)$. It is important to stress that these two
functions are completely unconstrained. For any choice of these two
functions one can in fact construct a Lagrangian $P(\phi,X)$, such
that the quadratic Lagrangian around the unperturbed solution has the
form (\ref{kessencefull}). Let us see explicitly how this works. First
of all, given $\rho_Q + p_Q$ (and possibly the other contributions to
the total energy density and pressure coming from other components)
one can find the two functions $\rho_Q(t)$ and $p_Q(t)$ solving the
Friedmann equations.\footnote{One has to solve the continuity equation
  for quintessence, $\dot \rho_Q + 3H(\rho_Q+p_Q) =0$, where
  $\rho_Q+p_Q$ is a known function of time, together with the
  Friedmann equation, $H^2 =(\rho_Q + \rho_{\rm rest})/3\Mp^2$ where
  $\rho_{\rm rest}$ includes all the additional sources of energy
  density in the Universe. These equations can be integrated up to a
  constant in the initial condition. This ambiguity corresponds to a
  shift in the cosmological constant which does not enter in the
  action for perturbations.} At this point it is easy to check that
the Lagrangian
\begin{equation}
  P(\phi, X) = \frac12 (p_Q-\rho_Q)(\phi) + \frac12 (\rho_Q + p_Q)(\phi) X +
  \frac12 M^4(\phi) (X-1)^2
\label{kessback}
\end{equation}
has the solution $\phi = t$, gives the requested pressure and energy
density as a function of time, and gives eq.~(\ref{kessencefull}) as
the quadratic action for perturbations. Note that the dimension of
$\phi$ is that of an inverse of a mass.  The coefficient $M^4$ is time
dependent and for quintessence we expect that it varies with a time
scale of order $H^{-1}$. Somewhere in this paper, to simplify the
calculations we take $M^4=$ const.

One advantage of the action (\ref{kessencefull}) is that the
coefficients of all terms are physically measurable quantities -- we
will see below that $M^4$ is related to the sound speed of
perturbations. This standard form of rewriting the action of
quintessence perturbations does not suffer from field redefinition
ambiguities. Indeed, there is an infinite number of physically
equivalent Lagrangians $P(\phi,X)$ related by field redefinitions
$\phi \to \tilde \phi(\phi)$ but they all give the same action
(\ref{kessencefull}). Note that for most purposes the explicit
construction of the action in terms of $\phi$ is irrelevant. Indeed,
one is free to choose {\em any} function $\rho_Q(a)$ to describe the
evolution of the quintessence energy density as a function of the
scale factor. Then the action (\ref{kessencefull}) will describe the
perturbations around this background. In particular, one can always make a
field redefinition such that $\phi=t$, as we did in
eq.~(\ref{kessback}).

Let us see what are the theoretical constraints that we can put on the
general form of the action (\ref{kessencefull}). A basic requirement
that we will impose on our theory is that it is not plagued by ghosts,
i.e. that its kinetic energy term is positive, \begin{equation} \rho_Q+p_Q+4M^4>0
\;.  \end{equation}The presence of a sector with the ``wrong'' sign of the
energy implies that the Hamiltonian is unbounded from below. If one
studies this sector alone, no pathology arises as the sign of the
energy is a matter of convention. However, quintessence is (at least)
gravitationally coupled to the rest of the world, so that there is the
danger of exchanging energy between a healthy sector and a
negative-energy one without bound. Classically this is not a problem
if quintessence perturbations are very small and remain in the linear
regime. Therefore, for a quintessence with negligible clustering (with
speed of sound $c_s \sim 1$), there is no obvious classical danger. At
the quantum level the situation is more pathological. The vacuum is
unstable to the spontaneous decay into positive and negative energy
states and the decay rate is UV divergent and
strictly infinite in Lorentz invariant theories~\cite{Cline:2003gs}.
Although it has been shown that it is possible to cut-off this
instability in a non-Lorentz invariant theory~\cite{Holdom:2004yx}, in
this paper we take a more conservative approach and forbid the
existence of ghosts.

If we set $M^4=0$ in the general action (\ref{kessencefull}) we reduce
to the case of a standard quintessence field with canonical kinetic
term $(\partial \phi)^2$. In this case, forbidding the ghost implies
$\rho_Q+p_Q>0$. Thus, as it is well known, a scalar field with minimal
kinetic term can violate the null energy condition, equivalent in a
cosmological setting to $\rho+p<0$, only if it is a ghost~\cite{Caldwell:1999ew}. In this simple case the speed of sound of
scalar fluctuations is $c_s^2=1$. When $M^4$ does not vanish the speed
of sound of fluctuations differs from unity \cite{Garriga:1999vw} and
reads 
\begin{equation} 
c_s^2 = \frac{\rho_Q+p_Q}{\rho_Q+p_Q + 4M^4} \;.  
\end{equation}
One can see that, for $\rho_Q+p_Q>0$, i.e.~for positive $c_s^2$,
$M^4<0$ implies that scalar perturbations propagate
super-luminally~\cite{Bonvin:2007mw,Babichev:2007dw}. This is
problematic in a theory with a Lorentz invariant UV
completion~\cite{Adams:2006sv}.

\begin{figure}[ht]
  \centering
  \includegraphics[scale=1]{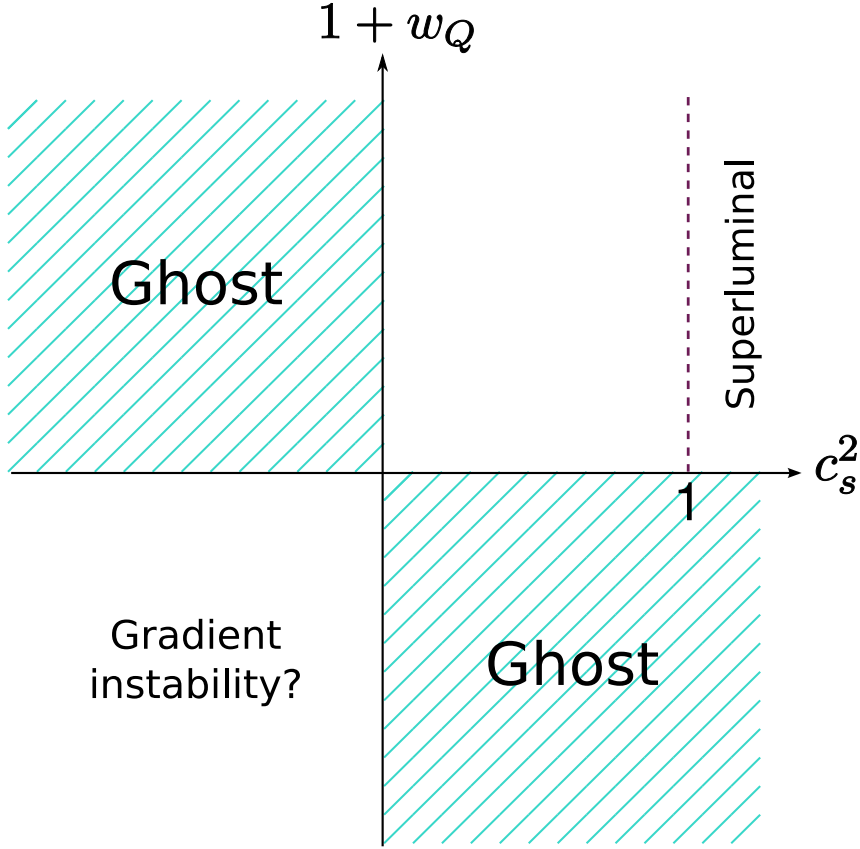}
  \caption{\small {\em The quintessential plane $1+w_Q$ vs.~$c_s^2$ in
      the case of $k$-essence. If we require the absence of ghosts,
      the sign of the spatial kinetic term is fixed to be the same as
      $1+w_Q$, so that one has to worry about gradient instabilities
      for $1+w_Q < 0$. For $1+w_Q>0$ one has superluminal propagation
      for $M^4 <0$.}}
  \label{fig:Qplaneghost}
\end{figure}

From the action (\ref{kessencefull}) we see that in the presence of
$M^4$ there is no generic connection between the violation of the null
energy condition and a wrong sign of the time-kinetic term. The
coefficient in front of $\dot \pi^2$ can be positive also when
$\rho_Q+p_Q<0$. On the other hand, $\rho_Q+p_Q$ fixes the sign of the
term in front of the spatial kinetic term $(\nabla \pi)^2$
\cite{Hsu:2004vr}. Thus, in absence of ghosts, the violation of the
null energy condition implies a negative speed of sound squared. The
constraints that we have derived are summarized in the {\em
  quintessential plane} $1+w_Q$ vs. $c_s^2$, represented in figure
\ref{fig:Qplaneghost}.

An imaginary speed of sound ($c_s^2<0$) represents a gradient
instability of the system. Taking $M^4 \gg |\rho_Q+p_Q|$ the gradient
term is suppressed and the instability is irrelevant for scales of
cosmological interest in the sense that the instability time is much
longer than the age of the Universe. Still, the instability is
relevant for short wavelengths so that it seems difficult to make
sense of the $w_Q< -1$ region~\cite{Vikman:2004dc}.  However, this conclusion is too hasty.
Indeed, when the term $(\nabla \pi)^2$ is suppressed, higher
derivative operators may become relevant as we will discuss in the
next section.

\subsection{Higher derivative terms and stability}
\label{sec:hd}
In the previous section we saw that in the limit in which the
quintessence speed of sound goes to zero, the standard spatial
kinetic term vanishes. Obviously the action (\ref{kaction}) is not the
end of the story: the full Lagrangian will contain higher derivative
operators such as $Q(X) R(\Box\phi)$ and these will give rise to the
leading higher derivative spatial kinetic term. A generic higher
derivative operator will however change the background solution of
(\ref{kaction}), while here we want to study the effect of higher
derivative operators on quintessence perturbations around a given
background, as we did in eq.~(\ref{kessback}). To keep the 
background unchanged, let us add
to the Lagrangian (\ref{kessback}) the operator
\begin{equation}
\label{Boxphi}
{\cal L}_{\bar M} = - \frac {\bar M^2}{2} (\Box\phi + 3 H(\phi))^2 \;.
\end{equation}
For reasons that will become clear later, we need
$\Mb^2>0$.\footnote{As $M^4$ also $\Mb^2$ can have a time dependence on a time scale of
  order $H^{-1}$. For simplicity, in the following we assume that it
  is constant.}
This term does not alter the background evolution $\phi=t$,
$\rho_Q(t)$ and $p_Q(t)$. 
Indeed, $\Box\phi +3 H(\phi)$ vanishes on the
background so that the operator is explicitly quadratic in the
perturbations. At quadratic order this operator reads
(neglecting for the moment metric perturbations)
\begin{equation}
\label{barM}
{\cal L}_{\bar M} = - \frac {\bar M^2}{2} \left(\ddot\pi +3 H \dot\pi -3 \dot H \pi
-\frac{\nabla^2\pi}{a^2} \right)^2 \;.
\end{equation}

One may worry about the presence of terms with higher time
derivatives, as these would na\"ively be associated with additional
solutions of the equation of motion. However, if one compares $\bar
M^2 \ddot \pi^2$ with the standard time kinetic term $M^4 \dot\pi^2$
of eq.~(\ref{kessencefull}) -- assuming $\bar M \sim M$ -- the former
is always suppressed with respect to the latter for frequencies below
the scale $M$. In general, we expect that for frequencies $\omega \sim
M$ all operators containing higher time derivatives become important,
so that the scale $M \sim \bar M$ sets the maximum energy scale for
which the theory makes sense: it is the energy cutoff. This is the
standard situation in an effective field theory: higher derivative
terms become important for energies of the order of the cutoff and at
lower energies they must be treated perturbatively. In particular,
there is no physical meaning in the new solutions that arise from
taking higher and higher time derivatives. We postpone a complete
discussion about this point to appendix~\ref{app:scaling}.  Notice
that the same argument cannot be used for the operator $-\bar
M^2(\nabla^2\pi)^2/2$. Indeed, in the limit of small $\rho_Q + p_Q$
there is no spatial kinetic term of the form $(\nabla \pi)^2$ so that
$-\bar M^2(\nabla^2\pi)^2/2$ is the leading spatial kinetic term. At
short scales we have a non-relativistic dispersion relation of the
form $\omega \simeq k^2/M$ which implies that energy and momentum
behave very differently (as we will see in appendix \ref{app:scaling}
they have different {\em scaling dimensions}). In particular, when
comparing the first and last terms in the brackets in eq.~(\ref{barM})
we have $\nabla^2\pi \sim M \dot \pi \gg \ddot \pi$, for energies
below the cutoff. This means that we can drop $\ddot\pi$ altogether
from eq.~(\ref{barM}).

It is important to stress that there is no fine tuning in the limit
$|\rho_Q+p_Q| \ll M^4$ -- or equivalently $|c_s^2|\ll 1$ -- as this
limit is ``technically natural'', i.e., there is a symmetry that is
recovered in the limit $\rho_Q+p_Q =0$. Indeed, as shown below, in
this limit we obtain the Ghost Condensate theory \cite{GhostCond},
which is invariant under the shift symmetry $\phi \to
\phi+\lambda$. In the presence of this symmetry the expansion of the
Universe drives the background solution to $\rho_Q+p_Q =0$. Thus,
models with very small speed of sound should be thought of as small
deformations of the Ghost Condensate limit \cite{Senatore:2004rj,StartingUniverse,EFTinflation}.

Let us come back to the issue of stability for $\rho_Q+p_Q<0$
including the new higher gradient term (\ref{barM}) and follow the
discussion in \cite{StartingUniverse}.  As we discussed in the
previous section, the only dangerous modes are those on scales much
smaller than the Hubble radius, as their instability rate can be
arbitrarily large; we thus concentrate on $k/a \gg H$. In this regime
the operator (\ref{barM}) further simplifies as the second and third
terms are negligible with respect to $\nabla^2\pi$. Considering the
action (\ref{kessencefull}) with the addition of the only remaining
operator $- \bar M^2 (\nabla^2\pi)^2/2$, the dispersion relation of
$\pi$ is thus modified to
\begin{equation}
  \left(\rho_Q + p_Q + 4 M^4\right)\omega^2 -(\rho_Q + p_Q)\frac{k^2}{a^2}
  - \Mb^2\frac{k^4}{a^4} = 0 \, .
  \label{dispersionrelation}
\end{equation}
For $\rho_Q + p_Q <0$ the $k^2$ gradient term has the unstable
sign, but in the presence of the new operator this instability is
confined to sufficiently large scales. In particular the fastest rate
of instability is given by
\begin{equation}
\omega_{\textrm{grad}}^2 \simeq -\frac{(\rho_Q + p_Q)^2}{\Mb^2 M^4}\, ,
\end{equation}
where we have taken $M^4 \gg |\rho_Q + p_Q|$. This gradient instability is
not dangerous when it is slower than the Hubble rate, i.e., when
\begin{equation}
\label{G_stability}
-\frac{\rho_Q + p_Q}{\Mb M^2} \lesssim H\;.
\end{equation}
It is clear that a larger $\Mb$ makes the gradient instability
slower.

However, a large $\Mb$ sources another form of instability, which
contrarily to the gradient instability is already present for $\rho_Q
+ p_Q = 0$ and was originally discussed for the Ghost Condensate
theory in \cite{GhostCond}. Indeed, when the coupling with gravity is
taken into account, the system shows a sort of Jeans instability
similarly to a standard fluid. To see this, let us introduce metric
perturbations and consider the limit $\rho_Q + p_Q = 0$. In this case,
the complete Lagrangian reads, neglecting the expansion of the
Universe,
\begin{equation}
  \label{GCaction}
  S = \!\int \!\ud^4 x \,
\Bigg[ 2 M^4 \dot{\pi}^2 - \frac {\bar M^2}{2} \left(\frac{\dot h}{2}
  -\nabla^2\pi\right)^2 \Bigg] \; .
\end{equation}
We see that a large $\Mb$ enhances the mixing of $\pi$ with gravity, 
i.e.~the Jeans instability.

Thus, the equation of motion for $\pi$ reads, in this case,
\begin{equation}
\label{pi_evolution_GC}
  \ddot{\pi} + \frac{\Mb^2}{4M^4} \nab^4 \pi
  = \frac{\Mb^2}{8M^4} \nab^2\dot{h}\,.
\end{equation}
The gravitational perturbation $h$ is sourced by the perturbations of $\pi$
through Einstein equations. In synchronous gauge, $h$
satisfies \cite{Ma:1995ey}
\begin{equation}
\label{h_evolution}
\ddot h = -\frac{1}{\Mp^2} (\delta \rho_Q+ 3 \delta p_Q)\;,
\end{equation}
where we have neglected the expansion of the Universe, and we have
introduced the reduced Planck mass $\Mp^2 \equiv (8 \pi G)^{-1}$.  The
stress-energy tensor can be straightforwardly derived by varying the
action (the complete expression of the stress-energy tensor is given
in appendix~\ref{app:SE}) and the leading term is $\delta \rho_Q =
4M^4 \dot \pi$ while the pressure perturbation is negligible.\footnote{Indeed, we have
\begin{eqnarray}
\delta \rho_Q &=& 4M^4 \dot \pi + \bar M^2 {\left(\frac{\ddot
      h}{2}-\nabla^2 \dot \pi \right)}\;, \\
\delta p_Q &=& \bar M^2 {\left(\frac{\ddot
      h}{2}-\nabla^2 \dot \pi \right)}\;.
\end{eqnarray}
For $k\ll M$, $\bar M^2\nabla^2 \dot \pi \ll M^4 \dot \pi$. Moreover,
eq.~(\ref{h_evolution}) shows that also the $\ddot h$ terms can be
neglected in front of $M^4 \dot \pi$, so that the operator
proportional to $\bar M^2$ gives a negligible contribution to the
stress-energy tensor.}
The solution of eq.~(\ref{h_evolution}) can be plugged back into
eq.~(\ref{pi_evolution_GC}). This yields the equation of motion of
$\pi$ taking into account its gravitational back-reaction,
\begin{equation}
  \ddot{\pi} + \frac{\Mb^2}{4M^4}\nab^4\pi
  = - \frac{\Mb^2}{2\Mp^2}\nab^2\pi \, .
\end{equation}
Mixing with gravity induces an unstable $k^2$ term in the
dispersion relation, similarly to the gradient instability
discussed above. We can compute again the
fastest instability rate,
\begin{equation}
  \omega_{\textrm{Jeans}}^2 \simeq -\left(\frac{\Mb M^2}{\Mpl^2}\right)^2 \, .
\end{equation}
As expected, in this case the instability gets worse for large $\bar
M^2$, i.e., when the mixing with gravity is enhanced. By imposing that
this instability rate is smaller than the Hubble rate\footnote{A more
  careful analysis \cite{ArkaniHamed:2005gu} indicates that this
  condition is very conservative and much larger instability rates can
  be experimentally allowed.} we obtain
\begin{equation}
  \label{J_stability}
  \frac{\Mb M^2}{\Mpl^2} \lesssim H \, .
\end{equation}
Requiring that both stability conditions~\eqref{G_stability}
and~\eqref{J_stability} are satisfied we get the
window~\cite{StartingUniverse}
\begin{equation}
  \label{stabilitywindow}
   -(1+w_Q) \Omega_Q \lesssim \frac{\Mb M^2}{H \Mpl^2} \lesssim 1 \, .
\end{equation}

In conclusion, considering higher derivative terms, a quintessence
model with $w_Q \le -1$ can be completely stable inside the window of
parameters~\eqref{stabilitywindow}. On the other hand,
eq.~\eqref{stabilitywindow} indicates that it is difficult to avoid
instabilities when $(1+ w_Q) \Omega_Q \ll - 1$.  These stability
constraints were already obtained, for $\Omega_Q = 1$,
in~\cite{StartingUniverse}.


\section{Phenomenology on the quintessential plane}
\label{sec:plane}

\subsection{$k$-essence vs.~Ghost Condensate}

Coming back to the quintessential plane of figure
\ref{fig:Qplaneghost}, in the previous section we have learned an
important lesson: {\em the gradient instabilities for $w_Q < -1$ can
  be made harmless by higher derivative operators.} Thus, part of the
lower left quadrant of the quintessential plane is allowed.

To discuss the phenomenology of these models (for a related
discussion see \cite{Mukohyama:2006be}) let us write the full
action for perturbations including the higher derivative operator
(\ref{Boxphi}):
\begin{eqnarray}
  \label{actionfull}
  S & = & \!\int \!\ud^4 x \, a^3
\Bigg[ \frac{1}{2} \left( \rho_Q+p_Q +4 M^4  \right) \dot{\pi}^2
  - \frac{1}{2}(\rho_Q+p_Q) \frac{(\nab \pi)^2}{a^2}
  + \frac{3}{2} \dot{H} (\rho_Q+p_Q)  \pi^2  -
\frac{1}{2}(\rho_Q + p_Q)\dot{h}\pi \nonumber \\ & & -\frac{\bar M^2}{2} \left(3 H \dot\pi -3 \dot H \pi
+ \frac{\dot h}{2} -\frac{\nabla^2\pi}{a^2} \right)^2\Bigg] \; .
\end{eqnarray}
First of all, note that it is not possible to switch off quintessence
perturbations for $\rho_Q+p_Q\neq 0$; doing it by hand would give
gauge dependent unphysical results. On the other hand, the converse is
not true: even for $\rho_Q+p_Q = 0$ perturbations may still be
present, as in the Ghost Condensate case.

We saw that the operator in the second line of eq.~\eqref{actionfull} allows the stabilization
of the short scale gradient instability; on the other hand, for
cosmological purposes we are interested in very large scales. Let us
see whether this operator is relevant for scales of the order of the
Hubble radius (although our discussion will extend to all scales of
cosmological interest). We want to show that when
\begin{equation}
|\rho_Q + p_Q| \gg \bar M^2 H^2 \label{KE_limit}
\end{equation}
the higher derivative operator can be neglected when discussing the
cosmological clustering of quintessence. In this case we reduce to a
standard $k$-essence model, with the only difference that there are no
short-scale instabilities even for $w_Q <-1$.  On the other hand in
the opposite case,
\begin{equation}
|\rho_Q + p_Q| \ll \bar M^2 H^2\;, \label{GC_limit}
\end{equation}
all the terms in the action (\ref{actionfull}) proportional to $\rho_Q
+ p_Q$ can be neglected. In this case the model reduces to the Ghost
Condensate theory.

Verifying the existence of these two regimes is quite
straightforward. For instance the dispersion relation at $k/a \sim H$
is dominated either by $(\rho_Q + p_Q) (\nabla\pi)^2$ or by $\bar M^2
(\nabla^2\pi)^2$ depending on the hierarchy between $|\rho_Q + p_Q|$
and $\bar M^2 H^2$. The same applies for the operators involving the
metric perturbation, $(\rho_Q + p_Q) \dot h \pi$ and $\bar M^2 \dot h
\nabla^2\pi$. This check can be done for all the other operators, by
taking $\nabla/a \sim H$ and considering that time derivatives are at
most of order $H$. The existence of these two regimes can also be seen
by looking at the stress-energy tensor (see appendix~\ref{app:SE}).

\begin{figure}[!!!ht]
  \centering
  \includegraphics[scale=1]{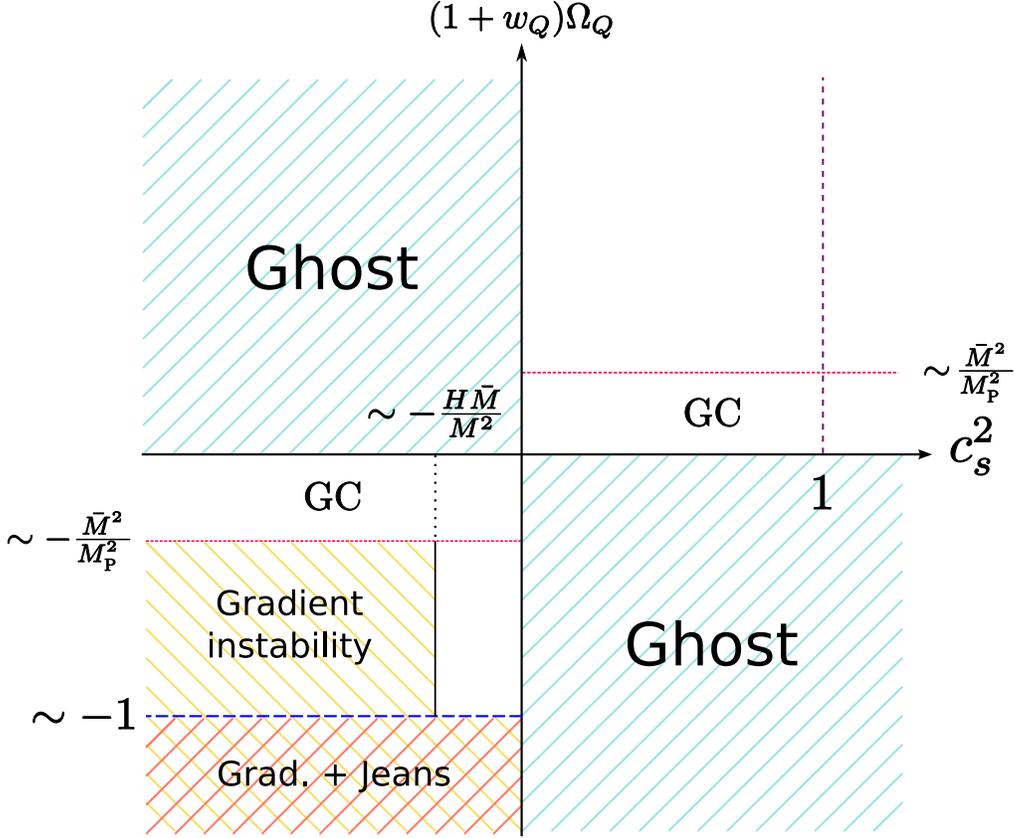}
  \caption{\small {\em On the quintessential plane, we show the
      theoretical constraints on the equation of state and speed of
      sound of quintessence, in the presence of the operator $\bar M$.
      Instability regions are dashed.  Where $1+w_Q$ and $c_s^2$ have
      opposite sign we have a ghost-like instability corresponding to
      negative kinetic energy.  For $w_Q<-1$, the dashed regions in
      the left-lower panel is unstable by gradient $(c_s^2 \lesssim -
      H \bar M /M^2)$ and Jeans $((1+w_Q)\Omega_Q \lesssim -1)$
      instabilities, while the strip close to the vertical axis
      corresponds to the stability window (\ref{stabilitywindow}).
      Furthermore, the strip around the horizontal axis given in
      eq.~(\ref{strip}) corresponds to the Ghost Condensate. Above
      this region, in the right-upper panel, we find standard
      $k$-essence.}}
  \label{fig:Qplane}
\end{figure}

Now we can go back and complete our quintessential plane. When $w_Q$
is close to $-1$,
\begin{equation}
\label{strip}
-\frac{\Mb^2}{\Mp^2} \lesssim (1+w_Q) \Omega_Q \lesssim
\frac{\Mb^2}{\Mp^2} \;,
\end{equation}
the model behaves as the Ghost Condensate. We can estimate the width
of this region by imposing the absence of Jeans instability,
eq.~(\ref{J_stability}). Assuming $M \sim \bar M$ one gets a rough
upper bound: $\bar M \lesssim 10\;{\rm MeV}$ \cite{GhostCond}. A more
accurate analysis shows that this limit is much too conservative and
it can be relaxed to $\bar M \lesssim 100\;{\rm GeV}$
\cite{ArkaniHamed:2005gu}.  Even in this case the window above is
extremely tiny:
\begin{equation}
  |1+w_Q| \Omega_Q \lesssim 10^{-34} \;. 
\end{equation}
We can therefore draw an important conclusion:
only for values of $w_Q$ which are observationally indistinguishable
from the cosmological constant, does quintessence behave as the Ghost
Condensate on cosmological scales. This regime corresponds to the
strip around the horizontal axis $w_Q=-1$ in figure
\ref{fig:Qplane}. Notice that in this region the dispersion relation
is of the form $\omega \propto k^2$, so that the speed of sound
$c_s^2$ is not well defined, i.e.~it becomes scale dependent.

On the other hand, for any value of $w_Q$ which is appreciably
different from the one of the cosmological constant, the model reduces
to $k$-essence as higher derivative terms are cosmologically
irrelevant.  Their only role is to stabilize the short scale gradient
instabilities for $w_Q<-1$.  Although in practice not relevant, note
however that $w_Q$ cannot be made arbitrarily negative.
This is shown by
eq.~(\ref{stabilitywindow}) and, in the quintessential plane, it
excludes the bottom shaded region of the lower-left quadrant.

Let us now constrain the values of the speed of sound
$c_s^2$. For $w_Q>-1$ there are no constraints, besides the possible
limit $c_s^2 \le 1$ already discussed. For $w_Q<-1$ the speed of sound
is negative and very small, as it is constrained by the absence of
gradient instability, eq.~(\ref{G_stability}),
\begin{equation}
- c_s^2 \simeq -\frac{\rho_Q +p_Q}{4M^4} \lesssim \frac{H \Mb }{M^2} \, .
\label{ciccio}
\end{equation}
We can numerically constrain the right-hand side of this equation by
considering that the scales $M \sim \bar M$ represent the cutoff of
our effective field theory. By requiring this cutoff to be larger than
the minimum scale at which gravity has been proved, i.e., $M \gtrsim
10^{-3}{\rm eV} $, and using in eq.~(\ref{ciccio}) the value of the
Hubble parameter today, $H_0 \sim (10^{-3}{\rm eV})^2 /\Mp$,
we obtain
\begin{equation}
  -c_s^2 \lesssim \left(\frac{H_0}{\Mpl}\right)^{1/2}  \sim 10^{-30}\, .
  \label{soundspeedbound}
\end{equation}
Thus, for all practical purposes the speed of sound can be taken to be
exactly zero. On the quintessential plane in figure \ref{fig:Qplane}, in the lower-left quadrant,
we can only live in a tiny strip along the vertical axis.
Notice however that there is no fine
tuning in keeping $c_s^2$ extremely small. Indeed, as we discussed, in the limit of
Ghost Condensate $c_s^2$ vanishes exactly for symmetry reasons. Thus, the
speed of sound remains small for tiny deviations from this limit.

\subsection{Including dark matter}
\label{sec:including_dm}

After the discussion about the stability constraints, we would like to
understand the dynamics of quintessence perturbations and their impact
on cosmological observations. In order to do this, we will now study
quintessence in the presence of cold dark matter, which
gravitationally sources quintessence perturbations. A thorough
analysis of the phenomenology of these models is beyond the scope of
this paper. Here we want to focus on the main qualitative features in
the various limits.

Let us start from the Ghost Condensate limit (\ref{GC_limit}). It is
known that the Ghost Condensate affects only short scales, i.e., $\pi$
perturbations induce a modification of the Newtonian potential at
scales parametrically smaller than the Hubble scale
\cite{GhostCond}. Therefore, we expect to have extremely small effects
on cosmological scales. To verify that this is the case, we can study
the action (\ref{actionfull}) in the limit of $\rho_Q+p_Q=0$. This reads
\begin{equation}
\label{actionGC}
  S =  \!\int \!\ud^4 x \, a^3
\Bigg[ 2 M^4 \dot{\pi}^2
 -\frac{\bar M^2}{2} \left(3 H \dot\pi -3 \dot H \pi
+ \frac{\dot h}{2} -\frac{\nabla^2\pi}{a^2} \right)^2\Bigg] \; .
\end{equation}

For simplicity, let us momentarily disregard the first two terms in
parentheses,
\begin{equation}
S = \int  \!\ud^4 x \, a^3\Bigg[2M^4\dot{\pi}^2 - \frac{\Mb^2}{2}\left(\frac{\dot{h}}{2} - \frac{\nabla^2 \pi}{a^2}\right)^2\Bigg] \, .
\end{equation}
Notice that this is the action used in the Ghost Condensate paper
\cite{GhostCond}. The equation of motion for the $\pi$ perturbations
is given by
\begin{equation}
\ddot{\pi} + 3H\dot{\pi} + \frac{\Mb^2}{4M^4}\frac{\nab^4\pi}{a^4} = \frac{\Mb^2}{8M^4}\frac{\nab^2\dot{h}}{a^2} \, .
\label{pi}
\end{equation}
The gradient term on the left hand side can be neglected on
cosmological scales. Indeed, the time derivatives will be at least of
order $\sim \Mb k^2/(a M)^2$, so that the friction term in the previous
equation will always dominate the gradient term for $k/a \sim H$. As
we want to show that Ghost Condensate perturbations remain small, we
assume that the dark matter dominates the perturbed Einstein
equations. The validity of this assumption can be checked {\em a posteriori}.

In a matter dominated Universe $\dot h = -2 \dot {\delta}_m$
\cite{Ma:1995ey}, with $\delta_m \equiv \delta \rho_m/\rho_m$, which,
using $\dot{\del}_m = H \del_m$ and the background Friedmann equation,
leads to
\begin{equation}
\label{doth}
\dot{h} = -\frac{2}{3H}\frac{\delta \rho_m}{\Mpl^2}\,.
\end{equation}
We can now replace this as the source of Ghost Condensate perturbations
on the right-hand side of eq.~(\ref{pi}). This yields, neglecting the gradient term,
\begin{equation}
\ddot{\pi} + 3H\dot{\pi} \simeq - \frac{\Mb^2}{12
  M^4\Mpl^2}\frac{\nabla^2 \del \rho_m }{a^2 H }  \,.
\end{equation}
If we now assume that the initial quintessence perturbations are small
so that the homogeneous solutions are sub-dominant, similarly to what
happens in standard quintessence \cite{Dave:2002mn}, this equation can be
solved to give
\begin{equation}
\dot{\pi} = -\frac{\Mb^2}{24 M^4 \Mpl^2}\frac{\nabla^2 \del
\rho_m }{a^2 H^2} \;.
\end{equation}

As we discussed in the previous section, the energy density and
pressure perturbations of the Ghost Condensate are dominated by the
$M^4$ operator so that $\delta\rho_Q \simeq 4 M^4 \dot\pi$ and $\delta
p_Q \simeq 0$. Thus, on cosmological scales,
\begin{equation}
\label{GCsmall}
\delta\rho_Q \sim  \frac{\Mb^2}{\Mpl^2} \del \rho_m \;.
\end{equation}
From the simple estimate of $\bar{M}$ below eq.~(\ref{strip}), we
conclude that quintessence perturbations are negligibly small with
respect to dark matter perturbations, $\delta\rho_Q \lesssim 10^{-34}
\delta \rho_m$. It is straightforward to generalize this analysis
including the two terms in parentheses of eq.~(\ref{actionGC})
previously neglected and verify that eq.~(\ref{GCsmall}) remains
valid. The conclusion of eq.~(\ref{GCsmall}) is quantitatively
consistent with the (small) modification of the Newton law derived in
\cite{GhostCond}, as one can check for example in their eq.~(7.11) for
$k/a \sim H$ and $\omega \sim H$.

Close to the $w_Q=-1$ line, we saw that there are no appreciable
effects of perturbations on cosmological scales which can help in
distinguishing quintessence from a cosmological constant; all the
interesting dynamics is limited to short scales. As we move away from
the $w_Q=-1$ line we enter in the $k$-essence regime, as we pointed
out in the previous section. The case $w_Q > -1$ is well studied in
the literature (see, for instance, \cite{Erickson:2001bq,DeDeo:2003te}). The case $w_Q <
-1$ is much less studied: here we have a negative speed of sound
squared that is so small -- see eq.~(\ref{soundspeedbound}) -- that
can be taken to be zero for all practical purposes. With a small speed
of sound we expect quintessence to cluster on scales shorter than the
Hubble radius driven by dark matter gravitational potential wells. To study
this, let us repeat the calculation we just did in the Ghost
Condensate case for a $k$-essence with $c_s^2=0$.

For simplicity, let us assume for the moment that $M$ is constant.
Varying the action (\ref{kessencefull}) we get the equation of motion
for $\pi$,
\begin{equation}
4 M^4 (\ddot \pi +3 H \dot\pi) - (\rho_Q + p_Q) \frac{\nabla^2}{a^2}
\pi - 3 \dot H (\rho_Q + p_Q) \pi=-\frac12 (\rho_Q + p_Q)\dot h \;.
\end{equation}
Small $|c_s^2|$ is equivalent to $|\rho_Q + p_Q| \ll M^4$ so that the
gradient and mass term in this equation can be neglected. Using again
$\dot h = -2 \dot\delta_m$ \cite{Ma:1995ey}, 
we thus have
\begin{equation}
4 M^4 (\ddot \pi +3 H \dot\pi) = (\rho_Q + p_Q)\dot \delta_m \;.
\end{equation}
One can verify that, neglecting decaying modes, the solution of
this equation is
\begin{equation}
\label{dQdm}
\delta_Q = \frac{1+w_Q}{1- 3 w_Q} \delta_m \;.
\end{equation}
It is easy to show that this equation holds, for constant $w_Q$, for a general time
dependent speed of sound which satisfies $|c_s^2(t)| \ll 1$.

Equation (\ref{dQdm}) describes quintessence perturbations
both for positive and negative $1+w_Q$. When $w_Q > -1$
quintessence energy density clusters in the dark matter potential wells, while
in the opposite case $w_Q < -1$ it escapes from them
\cite{Weller:2003hw}. However, clustering of quintessence remains small
compared to dark matter as the coupling with gravity is suppressed by
$1+w_Q$. For very small values $|1+w_Q|\sim \Mb^2/\Mp^2$ we smoothly enter in
the Ghost Condensate regime. Indeed it is easy to see that
eq.~(\ref{dQdm}) smoothly 
matches eq.~(\ref{GCsmall}) in the intermediate regime.

It is important to stress that for $w_Q < -1$ the speed of sound is
constrained to be so small that quintessence effectively clusters on
all scales. It would be interesting to understand the effect of the
short scale clustering on structure formation. We will come back to this point in the conclusion, section \ref{sec:conclusions}.

In this section we have studied the phenomenology of quintessence in
various regimes of $1+w_Q$. Quintessence perturbations smoothly turn
off when we approach the cosmological constant limit $w_Q = -1$ from
both sides. This suggests that in general there is no pathology in
crossing the $w_Q = -1$ line, as we discuss in the next section.


\section{Crossing the phantom divide}
\label{sec:divide}

It has been claimed that, during its evolution, single field
quintessence cannot cross the $w_Q = -1$ line as perturbations become
pathological. For this reason this line has been dubbed ``phantom
divide'' \cite{HuPhantom}. However, there is no real pathology in
crossing this line, besides the fact that for $w_Q<-1$ short-scale
gradient instabilities must be stabilized
\cite{Vikman:2004dc,CaldwellDoran}.  If one does not take into account
higher derivative terms, a negative $c_s^2$ leads to catastrophic
instabilities at short scales. Once instabilities are cured as we
discussed in the previous sections, crossing the phantom divide
becomes trivial.\footnote{Models that cross the phantom divide have
  been found in $f(R)$ theories of gravity. However, for $w_Q <-1$ these
  models are equivalent to scalar fields with negative kinetic energy,
  i.e. to ghosts. Thus, according to our stability constraints, they
  are forbidden.}

Indeed, the Lagrangian (\ref{kessback}) gives an explicit way to
construct a model which crosses the phantom divide. If one assumes for
simplicity that quintessence is the only component in the
Universe,\footnote{The conclusions drawn in this section also hold when
  including dark matter.} the crossing of the phantom divide
corresponds to a change of sign of $\dot H$. In particular,
considering only quintessence, one can use Friedmann equations to
recast the Lagrangian (\ref{kessback}) in the form (higher derivative
operators will be considered later)
\begin{equation}
P(X,\phi) = - 3 \Mp^2 H^2(\phi) - \Mp^2 \dot H (\phi) (X+1) + \frac12
M^4(\phi) (X-1)^2 \;.
\label{kessbackH}
\end{equation}
This is similar to what happens in inflation, where the inflaton is
the only relevant component in the Universe \cite{EFTinflation}.

\begin{figure}
\includegraphics[width=8cm]{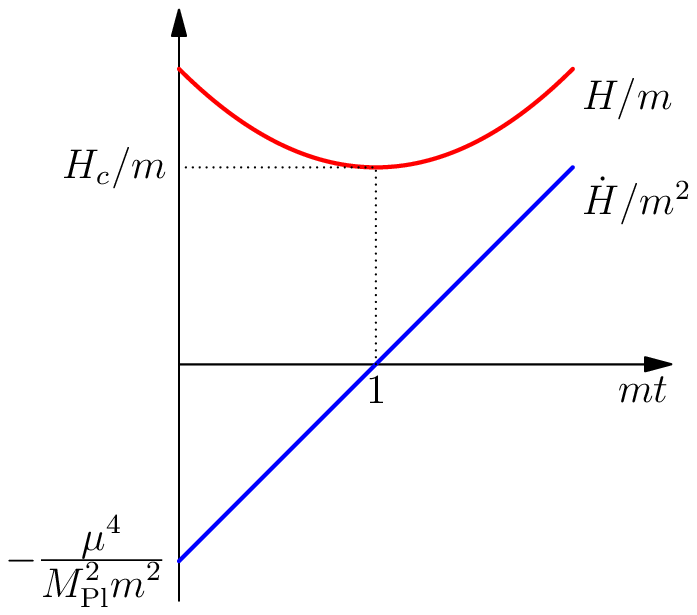}
\hfill
\includegraphics[scale=1]{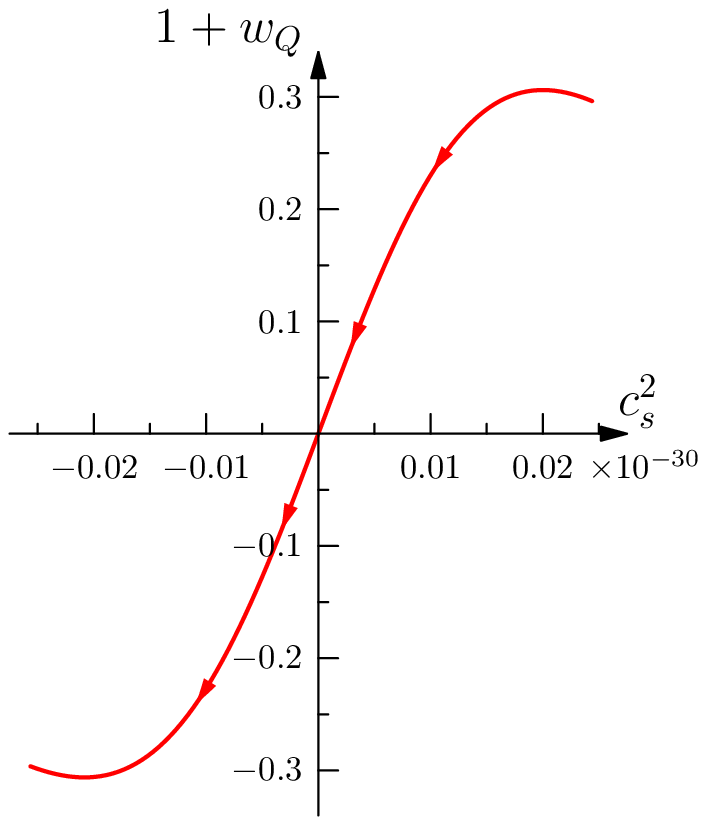}
\caption{\small\it Example of phantom divide crossing, as given by
  eqs.~(\ref{Hdt}) and (\ref{Ht}), where we have defined $H_c\equiv
  H_* - \mu^4/(2m\Mp^2)$.  Left figure: behavior of $H$ and $\dot H$;
  the crossing of $w_Q=-1$ takes place at $t=m^{-1}$ when $\dot H=0$
  and $H=H_c$. Right figure: trajectory on the quintessential plane. }
\label{fig3}
\end{figure}

As an example, we consider the case where $\dot H$ evolves linearly in
time and changes sign from negative to positive,
\begin{equation}
\label{Hdt}
\dot H (t) = \frac{\mu^4}{\Mp^2} (mt - 1) \;.
\end{equation}
This implies that $H(t)$ will be a parabola of the form 
\begin{equation}
\label{Ht}
H(t) = \frac{\mu^4}{\Mp^2} \left(\frac{m}{2} t^2 - t\right) + H_* \;,
\end{equation}
as shown in figure~\ref{fig3} (left panel).
Using the general expression (\ref{kessbackH}), we deduce that the
Lagrangian
\begin{equation}
P(\phi, X) = - 3
  \left[\frac{\mu^4}{\Mp}\left(\frac{m}{2} \phi^2 -\phi \right)+ \Mp
    H_*\right]^2  + \mu^4 (m\phi - 1) \left[(\partial\phi)^2-1\right]+
\frac12 M^4(\phi) \left[(\partial\phi)^2+1\right]^2
\end{equation}
admits the background solution $\phi = t$ and the cosmological evolution
(\ref{Hdt}) and (\ref{Ht}). Note that there are no theoretical limitations on
the choice of the background evolution $H(t)$. Indeed, we can cross
the phantom divide as many times as we want. For example, choosing
$\dot H (t) \propto \sin(m t)$ the cosmological evolution keeps
oscillating up and down around $w_Q=-1$!

We can now study perturbations around a solution crossing $w_Q=-1$ to
show that no pathology arises. The evolution equation derived from the
action (\ref{kessencefull}) reads
\begin{equation}
\label{nohigher}
  \left(\rho_Q + p_Q + 4 M^4 \right) \ddot{\pi}
  + \frac{1}{a^3} \de_t \left[ a^3 (\rho_Q + p_Q + 4 M^4) \right] \dot {\pi}
 - 3 \dot{H} (\rho_Q + p_Q) \pi  - (\rho_Q + p_Q) \frac{\nab^2
   \pi}{a^2} = - \frac{1}{2} (\rho_Q + p_Q) \dot{h} \, .
\end{equation}
At the phantom divide, the last three terms of this equation vanish
but the equation is clearly non-singular. In our approach, it is
manifest that the speed of sound squared changes sign at $w_Q=-1$. In the
example above $c_s^2$ is given by
\begin{equation}
c_s^2 = \frac{\mu^4(1 - m t)}{ \mu^4(1 - m t) + 2M^4 }\, ,
\end{equation}
and the trajectory of the crossing on the quintessential plane is
shown in figure~\ref{fig3} (right panel).  The stability in the
$w_Q<-1$ region requires that $|c_s^2|$ remains extremely small so
that it is mandatory to have a hierarchy between $\mu$ and $M$, $\mu
\ll M$. As we discussed, this hierarchy is naturally realized.

Another approach to study perturbations is using a fluid description,
as $k$-essence is equivalent to a perfect fluid.  In synchronous gauge, denoting with a prime the derivative with respect 
to the conformal time $d \eta \equiv dt/a$ and defining ${\cal H}=a H$, 
the fluid equations read, in Fourier space, (see for example \cite{BeanDore})
\begin{align}
  &\del_Q' + 3 \mathcal{H} (c_s^2 - w_Q) \delta_Q = - (1 + w_Q) \left[ k^2
    + 9 \mathcal{H}^2 (c_s^2 - c_a^2) \right] \frac{\th_Q}{k^2}
  - (1 + w_Q) \frac{h'}{2} \;, \label{continuity}\\
  &\frac{\th_Q'}{k^2} + \mathcal{H} (1 - 3 c_s^2) \frac{\th_Q}{k^2} =
  \frac{c_s^2}{1 + w_Q} \del_Q \;, \label{euler}
\end{align}
where $\theta_Q$ is the divergence of the velocity field of
quintessence, $\theta_Q \equiv i k^i T^0_{\ i}/ (\rho_Q +p_Q)$
\cite{Ma:1995ey}, $c_s^2$ is $\delta p_Q/\delta \rho_Q$ calculated in
a velocity orthogonal gauge ($T^0_i =0)$ \cite{Kodama:1985bj} and it corresponds to the speed of sound squared that
can be deduced from the $\pi$ Lagrangian\footnote{The velocity
    orthogonal condition $T^0_i =0$ is equivalent to the condition
    $\pi =0$. As $\phi$ is unperturbed in this gauge, the perturbations of pressure
    and energy density only come from fluctuations of $X$, i.e.~$\delta p_Q/\delta \rho_Q =
    p_Q,_X/\rho_Q,_X = P_X/(2 P_{XX} X + P_X)$, which is the speed of
    sound $c_s^2$ which appears in the $\pi$ Lagrangian (\ref{kessence}).}, to be distinguished from the
adiabatic speed of sound squared, $c_a^2 \equiv
{\dot{p}_Q}/{\dot{\rho}_Q} = w_Q - {\dot{w}_Q}/({3 H (1 + w_Q)})$. The
absence of pathologies at $w_Q\to -1$ can also be shown in this
formalism.  Indeed, in the continuity equation the divergence of
$c_a^2$ is compensated by the prefactor in front of the squared
brackets, while the $1+w_Q$ term at the denominator in the Euler
equation is harmless as $c_s^2$ also vanishes for $w_Q\to -1$. Thus
both $\delta_Q$ and $\theta_Q$ are continuous through the divide. This
is not surprising as $\theta_Q$ is just the Laplacian of the scalar
perturbation $\pi$, $\pi = a \,\theta_Q/k^2$.

At this point the reader may be puzzled: in the previous sections we
stressed that close to the $w_Q = -1$ line quintessence behaves like
the Ghost Condensate on cosmological scales, while
eq.~(\ref{nohigher}) as well as the fluid eqs.~(\ref{continuity}) and
(\ref{euler}) do not contain higher derivative terms. Let us see why
these additional terms are irrelevant in realistic cases of crossing
the phantom divide.  With these new terms, the equation of motion for
$\pi$ derived from the full action (\ref{actionfull}) is obviously
still continuous so that also in this case the line $w_Q = -1$ can be
crossed smoothly.  The operator proportional to $\Mb^2$ dominates in
the Ghost Condensate strip around the $w_Q = - 1$ line. However, this
happens only in the extremely narrow range $|1+w_Q| \lesssim
\Mb^2/\Mp^2 \lesssim 10^{-34}$. The equation of state parameter $w_Q$
will stay in this range only for a time much smaller than $H^{-1}$,
unless its evolution is tremendously slow. Thus $\pi$ has no time to
evolve in the Ghost Condensate regime, so that for all practical
purposes one can totally neglect this strip around the $w_Q = -1$ line
on cosmological scales.

We have seen that $k$-essence can be described with the fluid
equations (\ref{continuity}) and (\ref{euler}). Even including higher
derivative terms, quintessence remains a perfect fluid (see appendix
\ref{app:SE}) but does not satisfy the fluid equations
(\ref{continuity}) and (\ref{euler}) as these assume a linear
dispersion relation. However, as we discussed, higher derivative terms are
phenomenologically irrelevant on cosmological scales, so that one can
still use the fluid description above when comparing with
observations.

From a practical point of view we conclude that, when comparing with
observations a dark energy model which crosses the phantom divide, it
is consistent and theoretically motivated to set $c_s^2=0$. On the
other hand, it is inconsistent to turn off perturbations as sometimes
done in the literature.



\section{Additional higher derivative operators}
\label{sec:addhigh}

As we discussed, higher derivative operators become relevant when the
speed of sound is very close to zero. This regime is particularily
interesting when $w_Q<-1$ so that in the following we will consider mostly
the case $\rho_Q + p_Q <0$.

Theories with very small $c_s^2$ should be thought of
as tiny deformations of the Ghost Condensate theory \cite{Senatore:2004rj,StartingUniverse,EFTinflation}: in this limit one
recovers the shift symmetry $\phi \to \phi + \lambda$, so that a small
deviation from the Ghost Condensate is technically natural. In the
Ghost Condensate limit there is an additional symmetry that one can
impose, i.e.~the parity symmetry $\phi \to - \phi$. The background
$\phi = t$ in Minkowski space breaks this parity symmetry and the time
reversal symmetry to the composition of the two; the theory of
perturbations is thus invariant under $\pi \to - \pi$, $t \to - t$
\cite{GhostCond}. This symmetry is present only when the background
metric is Minkowski: in de Sitter there is a preferred time direction
singled out by the expansion. In this case terms violating the
symmetry will be proportional to $H$, and thus tipically suppressed by
$H/M$. In this paper we have considered small departures from the
Ghost Condensate limit, i.e., tiny breakings of the shift symmetry.
These also generate terms which are not invariant under parity $\pi
\to -\pi$, $t \to -t$, as for instance the mass term in
eq.~(\ref{kessencefull}).
These terms
will be of the same order of magnitude, i.e.~suppressed by $H/M$ as we
are assuming that $H^{-1}$ is the typical time scale of evolution of the
operators.

However, one can also consider the case when the parity symmetry $\phi
\to - \phi$ is absent in the Ghost Condensate
limit~\cite{StartingUniverse}. This happens for example if we add to
the $k$-essence Lagrangian (\ref{kessback}) the operator
\begin{equation}
\label{Mhat}
{\cal L}_{\hat M} = -\frac{\hat M^3}{2} (\Box\phi+3 H)(X-1) \;,
\end{equation}
which again does not change the background evolution as it starts
quadratic in the perturbations.  For simplicity we assume that $\bar
M= 0$ and a constant $\hat M$.\footnote{ This assumption can be
  relaxed by having a time dependence with time scale of order
  $H^{-1}$, as in the case of $\Mb^2$.}  In synchronous gauge at
quadratic order this operator is
\begin{equation} {\cal L}_{\hat M} =\hat M^3 \dot\pi \left(\ddot \pi +3
H \dot\pi - 3\dot H \pi 
-\frac{\nabla^2\pi}{a^2} + \frac{\dot h}{2}\right ) \;.  
\end{equation}
The first two terms in the parentheses contribute (after an
integration by parts) to the time kinetic term. Assuming $\hat M \sim
M$ they can be neglected in comparison with $2 M^4 \dot\pi^2$. The
third term gives a mass term that is parametrically smaller than $H$
and can thus be neglected.

To discuss the stability and phenomenology of this model, let us write
the full action for perturbations, assuming $|\rho_Q+p_Q| \ll M^4$,
\begin{equation}
  \label{actionfull_hat}
  S  =  \!\int \!\ud^4 x \, a^3
\Bigg[ 2 M^4  \dot{\pi}^2
  - \frac{1}{2}(\rho_Q+p_Q) \frac{(\nab \pi)^2}{a^2}
-\frac{1}{2}(\rho_Q + p_Q)\dot{h}\pi 
+ \hat M^3
\dot\pi \left( \frac{\dot h}{2} -\frac{\nabla^2\pi}{a^2}
  \right ) \Bigg] \; ,
\end{equation}
where we have neglected the mass terms.  This equation is analogous to
eq.~(\ref{actionfull}) for the $\bar M$ operator
(\ref{Boxphi}). Analogously to what we have done in
sections~\ref{sec:hd} and \ref{sec:including_dm} for the operator
proportional to $\bar M^2$, we will now study the stability and
phenomenology on cosmological scales with the operator ${\cal L}_{\hat
  M}$. In appendix \ref{app:modgrav} we briefly study the effect of
this operator at short distances, i.e. the modification of gravity
induced by it.


\subsection{Stability constraints with $\hat M$}


Let us first study the stability of the system neglecting other
sources of gravity. The equation of motion for $\pi$ derived varying
(\ref{actionfull_hat}) reads
\begin{equation}
\ddot\pi +3 H \dot\pi -  \frac{\rho_Q + p_Q}{4M^4} \frac{\nabla^2\pi}{a^2} -
\frac{\hat M^3 H}{4 M^4} \frac{\nabla^2\pi}{a^2} 
= - \frac{\rho_Q+p_Q}{8M^4} \dot h  - \frac{\hat M^3}{8 M^4} (\ddot h +3 H \dot
h) \;.\label{Mhat_pi}
\end{equation}
Notice that the operator $\hat M^3 \dot\pi \nabla^2\pi/a^2$ induces a
spatial kinetic term for $\pi$ proportional to $H$. Indeed, this
operator is a total derivative in Minkowski spacetime.  Choosing $\hat
M >0$, the spatial kinetic term has the ``healthy'' sign and can be
chosen sufficiently large to cure the gradient instability associated
to $\rho_Q + p_Q < 0$, giving a positive and very small $c_s^2$. This
also allows us to neglect the first term on the right hand side in
eq.~(\ref{Mhat_pi}).  To complete the stability analysis one has to
take into account the mixing with gravity, i.e., solve for $h$ in
terms of the quintessence stress-energy tensor using the Einstein
equation (\ref{h_evolution}), and plug the result back in the right
hand side of eq.~(\ref{Mhat_pi}).  This, similarly to what happens for
the Ghost Condensate, will give rise to a Jeans-like instability.

The contribution to the stress-energy tensor of the operator
(\ref{Mhat}) is (see appendix~\ref{app:SE})
\begin{eqnarray}
\delta\rho_Q & \supset & \hat M^3 \left(\frac{\dot h}{2} -
  \frac{\nabla^2\pi}{a^2}\right) \;, \label{drho_Mhat}\\
\delta p_Q & \supset & -2 \hat M^3 \left(\ddot\pi + 3 H \dot\pi\right) \;.
\end{eqnarray}
Given the small speed of sound, time derivatives are much smaller than
the spatial ones and the pressure perturbation is negligible, $\delta
p_Q \ll \delta \rho_Q$.  Concentrating on frequencies much larger than
the Hubble rate one can neglect the terms containing $H\dot \pi$ and
$H \dot h$ in eq.~(\ref{Mhat_pi}). A further simplification comes from
disregarding the standard $k$-essence contribution to the energy
density perturbation $\delta \rho_Q$, i.e.~$4M^4 \dot \pi$, in comparison
with $\hat M^3 \nabla^2 \pi/a^2$. Indeed, from eq.~(\ref{Mhat_pi}) we
have $\hat M^3 \nabla^2 \pi/a^2 \sim M^4 \ddot \pi/ H \gg M^4 \dot
\pi$.  Moreover, as we will see, the absence of Jeans instability will
impose $\hat M^3 \lesssim \Mp^2 H $. This implies that the term with
$\dot h$ in eq.~(\ref{drho_Mhat}) is negligible with respect to $\ddot
h$ in the Einstein equation (\ref{h_evolution}), that becomes 
\begin{equation}
\ddot h = \frac{\hat M^3}{\Mp^2} \frac{\nabla^2 \pi}{a^2}.  
\end{equation}
Plugging this into the right hand side of eq.~(\ref{Mhat_pi}) we
finally find
\begin{equation} \ddot\pi - \left( \frac{\rho_Q + p_Q}{4M^4} +
  \frac{\hat M^3 H}{4 M^4} - \frac{\hat M^6}{8 M^4 \Mp^2 }\right)
\frac{\nabla^2\pi}{a^2} = 0\;.\label{Mhat_pi2} 
\end{equation}
This same result would have been found using a more rigorous approach,
as done in~\cite{StartingUniverse}.  Again, as in the Ghost Condensate
case, mixing with gravity induces a Jeans-like instability,
represented by the last term in this equation.  Thus, for
$\rho_Q+p_Q<0$ we need to cure both the gradient and the Jeans
instabilities. This is possible for
 \begin{equation}
   - (1+w_Q) \Omega_Q \lesssim \frac{\hat M^3 }{\Mp^2 H} \lesssim  1\;. \label{sw_hat} 
\end{equation}
This stability window \cite{StartingUniverse} is analogous
to the one discussed in the Ghost Condensate case,
eq.~(\ref{stabilitywindow}).

We conclude that with the inclusion of the operator ${\cal L}_{\hat
  M}$ we can have a dispersion relation $\omega \propto k$ with
positive speed of sound squared; thus, there is no sign of instability even
for $\rho_Q+p_Q<0$.


\subsection{Including dark matter}


Analogously to what done in section~\ref{sec:including_dm}, to study
the phenomenology induced by the $\hat M$ operator we study
quintessence perturbations generated by the coupling with dark matter.
For simplicity we assume matter dominance. The action
(\ref{actionfull_hat}) gives the equation of motion for $\pi$ sourced
by the dark matter perturbation $\delta_m$,
\begin{equation}
\ddot \pi + 3H \dot \pi -\frac{H \hat M^3}{4 M^4} \frac{\nabla^2
  \pi}{a^2} = \frac{5}{8} \frac{\hat M^3}{M^4}H^2 \delta_m+ \frac{\rho_Q + p_Q}{4
M^4} H \delta_m\;, \label{eom_hat}
\end{equation}
where we have used $\dot h =-2 H \delta_m$ and neglected the gradient
term proportional to $\rho_Q+p_Q$ that is subdominant.  Since the
speed of sound $c_s^2 = H \hat M^3/(4 M^4)$ is very small and we are
interested in cosmological scales, one would na\"{i}vely neglect the
term with $\nabla^2 \pi$. This gives the solution
\begin{equation}
\dot \pi = \frac{H \hat M^3}{4 M^4} \delta_m + \frac{\rho_Q}{4
  M^4} \frac{1+w_Q }{1-3w_Q}   \delta_m\;.
\end{equation}
However, the approximation of neglecting the gradients is not good.
Indeed, when we plug this expression into the
energy density perturbation
\begin{equation}
\delta \rho_Q = 4 M^4 \dot \pi  + \hat M^3 \left(\frac{\dot h}{2} -
  \frac{\nabla^2 \pi}{a^2} \right)\;,
\end{equation}
there is a cancellation of terms proportional to $\hat M^3$ up to
gradient terms. Thus one is forced to go back to the equation of
motion (\ref{eom_hat}) and keep the term proportional to the
speed of sound.

Once we do that, we obtain
\begin{equation}
\delta_Q = - \frac{\hat M^6}{24 M^4 \Mp^2 \Omega_Q} \frac{\nabla^2}{H^2 a^2}
\delta_m + \frac{1+w_Q}{1-3w_Q} \delta_m\;. \label{drho_hat}
\end{equation}
This equation displays the existence of two different regimes, in
strict analogy with what happens in the Ghost Condensate case. For large
enough $|1+w_Q|$, the second term on the right hand side 
dominates and one recovers eq.~(\ref{dQdm}),
where the system behaves as standard $k$-essence. Note however that
even for $\rho_Q+p_Q <0$ there are no stability problems in the stability window (\ref{sw_hat}).
The dynamics of the system is dominated by the $\hat M$ operator only
when we are very close to $w_Q =-1$, i.e., for
\begin{equation}
- \frac{\hat M^6}{M^4 \Mp^2}\lesssim (1+w_Q) \Omega_Q\lesssim \frac{\hat M^6}{M^4 \Mp^2}\;. \label{stripMhat}
\end{equation}
This region is the analog of the Ghost Condensate strip around the
horizontal axis of figure \ref{fig:Qplane}.  In this range the first
term in eq.~(\ref{drho_hat}) dominates and $\delta_Q$ remains
extremely small with respect to $\delta_m$.  Although the dispersion
relation is of the form $\omega \propto c_s k$, quintessence does not
follow the simple fluid equations (\ref{continuity}) and (\ref{euler})
because of the presence of higher derivative operators. However, as in
the Ghost Condensate case, given the narrowness of the strip
(\ref{stripMhat}), for all practical purposes we can always use the
fluid equations with $c_s^2=0$, even when crossing the phantom divide.
In conclusion, the addition of the operator $\hat M$ can stabilize
$k$-essence in the phantom case $w_Q < -1$, and the phenomenology of
the model is the same as for a $k$-essence with $c_s^2=0$. This
general conclusion will hold even when considering both operators
$\bar M$ and $\hat M$ at the same time, and can be extended to all the
possible higher derivative operators, included in the general action
in appendix~\ref{app:unitary}.


\section{Conclusion and outlook}
\label{sec:conclusions}

In this paper we have studied the most generic action describing the
perturbations of a single field dark energy -- here called
quintessence -- around a given background. We have constructed the
action by adding to the $k$-essence Lagrangian higher derivative
operators that leave the background evolution invariant. Using this
action, we have reproduced the results of \cite{StartingUniverse}
concerning the theoretical constraints on the equation of state
parameter $w_Q$ as a function of the speed of sound squared $c_s^2$,
by the requirement that perturbations are ghost-free -- i.e., that
their kinetic energy is positive -- and that there are no
gradient-like instabilities. These constraints have been conveniently
represented on the quintessential plane $(1+w_Q)\Omega_Q $
vs.~$c_s^2$, in figures~\ref{fig:Qplaneghost} and \ref{fig:Qplane}.

In particular we have considered the case $w_Q<-1$, which is commonly
believed to be unstable, and we have shown that for very small $c_s^2$
both the gradient and the Jeans instabilities can be avoided and
perturbations stabilized \cite{StartingUniverse}. Higher derivative operators are crucial for
the stabilization.  Indeed, it is important to stress that taking an
extremely small $c_s^2$ does not represent a fine tuning, as in the
limit $c_s^2 \to 0$ we recover the Ghost Condensate theory which is
protected by the shift symmetry $\phi\to \phi+\lambda$. Thus, for
$w_Q<-1$ quintessence should be thought of as a small deformation of
the Ghost Condensate limit \cite{Senatore:2004rj,StartingUniverse,EFTinflation}.  When the higher order terms containing
$k^4$ dominate over the spatial kinetic term $c_s^2 k^2$ the
phenomenology reduces to that of the Ghost Condensate. This always
happens on small scales, where the higher order gradients must
dominate to stabilize the perturbations, but on cosmological scales
this only occurs for values of $w_Q$ extremely close to the one of the
cosmological constant, i.e., for $|1+w_Q|\Omega_Q \lesssim
10^{-34}$. Away from this tiny strip -- i.e., for all practical
purposes -- the behavior on cosmological scales is very different from
that of the Ghost Condensate: higher derivative terms are irrelevant
so that the phenomenology of the $w_Q<-1$ side of quintessence reduces
to that of a $k$-essence fluid with $c_s^2 =0$.

Furthermore, we have studied the behavior of quintessence
perturbations when crossing the so-called phantom divide $w_Q=-1$.  By
restricting the analysis to $k$-essence perturbations around a given
background crossing the phantom divide, we have shown that, as the
speed of sound vanishes exactly at the divide, perturbations remain
finite during the crossing. For $w_Q<-1$, higher derivative terms,
while irrelevant on cosmological scales, are essential to
stabilize the short scale gradient instabilities. We conclude that no
pathology arises during the crossing: the phantom divide can be
crossed without the addition of new degrees of freedom. We have illustrated this with an
example shown in figure \ref{fig3}. An important thing to retain is
that a consistent and theoretically motivated way of comparing with
data a dark energy evolution which crosses the phantom divide is to
set to zero the speed of sound of perturbations.

Our study motivates the possibility that quintessence has a virtually
vanishing speed of sound, especially when $w_Q < -1$.  Such a
quintessence can be detected through its effects on structure
formation. The speed of sound defines the sound horizon $\ell_{Q }
\equiv a \int c_s dt/a $, which sets the characteristic length scale
of smoothness of the perturbations. In the matter dominated era
$\ell_{Q } = 2 c_s H_0^{-1}$ for a constant $c_s$. Hence, for $c_s=1$ --
corresponding to the speed of sound of a scalar field with a canonical
kinetic term -- quintessence can cluster only on scales larger than
the Hubble radius while for $c_s=0$ it clusters on all scales, thus
affecting the gravitational potential and the formation of structures
of dark matter and galaxies.  The effect of a clustering quintessence
can be measured with the cosmic microwave background
\cite{DeDeo:2003te,Weller:2003hw,BeanDore,Hannestad:2005ak}, galaxy
redshift surveys \cite{Takada:2006xs}, large neutral hydrogen surveys
\cite{TorresRodriguez:2007mk}, or by cross-correlating the integrated
Sachs-Wolfe effect in the cosmic microwave background with large scale
structures \cite{Hu:2004yd,Corasaniti:2005pq}. For instance, in
\cite{Hu:2004yd,Takada:2006xs} it was found that with future surveys
it will be possible to measure a zero speed of sound of dark energy if
$w_Q \gtrsim -0.95$. As these analysis were restricted to positive
values of $1+w_Q$ only, it would be interesting to repeat them for
negative values. Notice that for a vanishing speed of sound, dark
energy will actively participate to the formation of non-linear
objects, affecting the halo bias. It would be interesting to evaluate
this effect, which gives additional signatures of these models.

Quintessence is perturbed by the presence of sources and thus modifies gravity as any other kind of matter. When
this modification happens on scales much smaller than the Hubble
radius one can properly talk about a theory of infrared modification
of gravity. This happens when quintessence is close to the Ghost
Condensate limit; in this case the modification of gravity is due to
the Jeans instability induced by higher derivative operators, and
persists even in Minkowski spacetime.  Modifications of gravity
induced by the Ghost Condensate have been studied in details in
\cite{GhostCond,ArkaniHamed:2005gu}. In this paper we have considered
also the additional operator proportional to $\hat M^3$
\cite{StartingUniverse}; it would be interesting to investigate the
deviation from General Relativity induced by this operator and its
possible observational consequences, extending the preliminary
analysis of appendix~\ref{app:modgrav}, where the treatment has been
restricted to linear perturbations in a Minkowski spacetime.

\section*{Acknowledgments}
It is a pleasure to thank Carlo Baccigalupi, Pier Stefano Corasaniti, Ruth Durrer, Steen
Hannestad, Wayne Hu, David Langlois, Alberto Nicolis, Uro\v{s} Seljak, and Martin Sloth for useful discussions and comments.
We are particularly in debt with Leonardo Senatore for 
collaboration in the early stages of the work.


\appendix


\section*{Appendix}
\section{Higher derivative operators in effective field theories}
\label{app:scaling}


In this appendix we want to study the Ghost Condensate and its
deformations from an effective field theory point of view. In
particular we want to show that, although the operator $(\nabla^2
\pi)^2$ dominates the dynamics,
operators containing higher time derivatives such as $\ddot \pi^2$
must be treated perturbatively.

In the Ghost Condensate limit, the free $\pi$ action is
\begin{equation}
S = \frac{M^4}{2} \int d^3 x dt \left[ \dot \pi^2 - \frac{(\nabla^2
  \pi)^2}{M^2} \right]\;, \label{GCaction_appendix}
\end{equation}
where we neglected the mixing with gravity -- as we are interested in
the high energy behavior of the theory -- and for simplicity we
assumed that there is a single scale $M$ ($M \simeq \Mb$).  This
action is manifestly invariant under the energy scaling
\cite{GhostCond}
\begin{equation}
E \to s E \;, \qquad   t \to s^{-1} t \;, \qquad x \to s^{-1/2} x\;,
\qquad \pi \to s^{1/4} \pi \;. \label{scaling_trans}
\end{equation}
As the theory is not Lorentz invariant, time and space behave
differently under rescaling, and $\pi$ does not scale as $s^1$ as in a
Lorentz invariant theory. (See for instance \cite{Polchinski:1992ed}
for an introduction to scaling in non-Lorentz invariant field
theories.)

What is the physical meaning of this scaling transformation? Assuming
that the theory is weakly coupled, the free action gives the leading
contribution to the correlation functions, so that these will be
invariant under the scaling above. For instance, a relativistic
massless scalar has scaling dimension~$1$. Thus, the two-point
function satisfies\footnote{Note that the scaling dimension has
  nothing to do with the mass dimension
of the field $\phi$. Indeed, eq.~(\ref{rel_phi}) remains the same if we
choose a non-conventional normalization of the action such that $\phi$
has not mass dimension $1$.}
\begin{equation}
\label{rel_phi}
\langle \phi \phi\rangle (x-y) = s^{-2} \langle \phi \phi\rangle \left(
  \frac{x-y}{s} \right) \qquad \Rightarrow \qquad
\langle \phi \phi\rangle \propto \frac{1}{|x-y|^2}\;.
\end{equation}
In the case of the action (\ref{GCaction_appendix}) above, the scaling
transformation (\ref{scaling_trans}) yields
\begin{equation}
\langle \pi \pi\rangle (\Delta t, \Delta \vec x) = s^{- 1/2}
\langle \pi \pi\rangle \left(\frac{\Delta t}{s}, \frac{\Delta \vec
  x}{s^{1/2}} \right) \;.
\end{equation}

Not only the scaling transformation gives information on the free
theory, but, more importantly, it allows one to estimate the effect of
different operators added to the free action. In particular, in the
Ghost Condensate case, one can check that all additional operators
allowed by symmetries have positive scaling dimensions, so that their
importance is suppressed by $E/M$ elevated to a positive power
\cite{GhostCond}. This implies that at low energy the theory is
perturbative. For instance, the leading irrelevant operator is $\dot
\pi (\nabla \pi)^2$, which has scaling dimension $1/4$.

Operators containing higher time derivatives
have positive scaling dimensions so that they must be treated
perturbatively. For instance $\ddot \pi^2$ has scaling dimension $2$,
so that at low energy it is negligible. Additional time derivatives
na\"{i}vely suggest the existence of more and more solutions of the
equations of motion. However, these solutions are non-perturbative in
the expansion parameter $E/M$, and there is no reason to expect that
they have any physical meaning. For example, taking seriously these solutions would
imply that the Minkowski vacuum is unstable when considering higher
order corrections to the Einstein-Hilbert action \cite{Simon:1990jn}.
The correct way of treating these terms is perturbatively, i.e.,
evaluating them using the lower order equations of motion
\cite{Simon:1990ic}. Following this logic,
the additional solutions studied in
the context of the Ghost Condensate theory in \cite{Kallosh07} are
non-physical, as already pointed out in \cite{Weinberg08}.

As we discussed in this paper, in certain regimes quintessence behaves
as a deformation of the Ghost Condensate theory. The free action
(\ref{GCaction_appendix}) is deformed by the addition of a $(\nabla
\pi)^2$ term,
\begin{equation}
S = \frac{M^4}{2} \int d^3 x dt \left[ \dot \pi^2  - c_s^2 (\nabla
\pi)^2 - \frac{(\nabla^2
  \pi)^2}{M^2} \right]\;. \label{mixed_action_appendix}
\end{equation}
In these cases the
dispersion relation is not exactly $\omega \sim k^2/M$ but it
contains also a linear term $\omega \sim c_s k $, with $c_s \ll
1$. (For this discussion we assume that $c_s^2$ is positive.)

\begin{figure}[!!ht]
  \centering
  \includegraphics[scale=0.8]{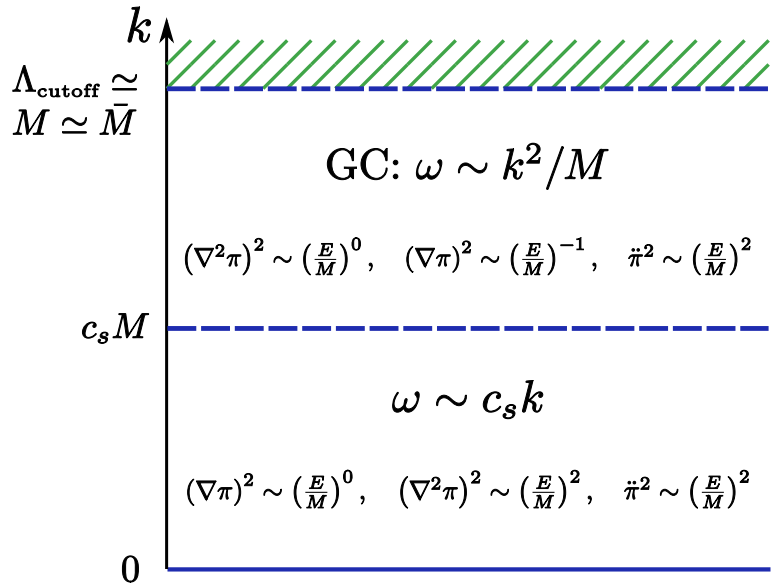}
  \caption{\small {\em The two scaling regimes as a function of the momentum $k$
    together with the scaling dimensions of some of the operators.
    Above: the Ghost Condensate regime where $\omega \sim k^2/M$; below:
  the regime where $\omega \sim c_s k$.}}
  \label{fig:scaling}
\end{figure}

The situation is now trickier than before because one cannot find a
scaling transformation which leaves the full action (\ref{mixed_action_appendix}) invariant.
On the other hand, one can separate two regimes, depending on which of
the gradient terms dominates, as illustrated in
figure~\ref{fig:scaling}.

For $k \gg c_s M$, the dispersion
relation is dominated by $(\nabla^2 \pi)^2$ and the theory behaves as
the Ghost Condensate. In this regime the scaling of all additional
operators can be obtained from eq.~(\ref{scaling_trans}). Notice
that now there is a relevant operator, $(\nabla \pi)^2$, that becomes
more and more important at low momenta (and energies). The coefficient
of this operator is however suppressed by the small deformation
parameter $c_s \ll 1$. Thus, it can be treated perturbatively as long
as $k \gg c_sM$.

On the other hand, when $k \ll c_s M$ the $(\nabla \pi)^2$
operator dominates the free action. In this regime the scaling becomes
the same as in the relativistic case. Now both $\ddot \pi^2$
and $(\nabla^2 \pi)^2$ are irrelevant operators with the same
scaling dimension $2$. However, time and spatial derivatives are still
on a different footing because the time derivatives are suppressed by
$c_s$ with respect to the spatial ones, $\omega \sim c_s k$. Thus
$\ddot \pi^2 \sim c_s^4 (\nabla^2 \pi)^2$ for $k \ll c_s M$.
In the intermediate regime $k \sim c_s M$ this suppression can also
be obtained from the Ghost Condensate limit. Indeed, at high momenta
the operator $\ddot \pi^2$ scales like $(k/M)^4$, so that it is
suppressed by $c_s^4$ for $k \sim c_s M$.

Even though these theories make perfect sense as effective field
theories, it is not clear whether one can find a UV completion.
In particular, the violation of the null energy condition may be
problematic in the context of black hole thermodynamics \cite{ArkaniHamed:2007ky}.


\section{Stress-energy tensor and fluid quantities}
\label{app:SE}

Here we compute the stress-energy tensor for the $k$-essence action
with the addition of the two higher derivative operators ${\cal
  L}_{\bar M}$ and ${\cal L}_{\hat M}$, i.e.,
\begin{equation}
S=\int \ud^4 x\,\sqrt{-g}\left[P(\phi,X)-\frac{\bar{M}^2}{2}(\square
\phi + 3H(\phi))^2 -\frac{\hat{M}^3}{2}(\square\phi + 3H(\phi))(X - 1)\right]\;.
\end{equation}
We start by computing, using eq.~\eqref{defTmunu} the
stress-energy contribution of the $k$-essence action, which can be written as
\begin{equation}
\label{ktensor}
(T_{\mu\nu})_P = 2 P_X (\phi,X)\partial_\mu\phi\partial_\nu\phi +
P(\phi,X)g_{\mu\nu} \, .
\end{equation}
It is well known that this stress-energy tensor can be put in the
perfect fluid form,
\begin{equation}
\label{perfectfluid}
T_{\mu\nu} = (\rho + p)u_\mu u_\nu + p g_{\mu\nu}\;,
\end{equation}
by defining the $k$-essence contribution to the
energy density and pressure respectively as
\begin{align}
\rho_P &\equiv 2XP_X - P \, ,\\
p_P &\equiv P \, ,
\end{align}
and the unit four-velocity of an observer comoving with the fluid as
\begin{equation}
u_\mu \equiv
\frac{\partial_\mu \phi}{\sqrt{-(\partial \phi)^2}} \label{velocity} \, .
\end{equation}
The stress-energy tensor (\ref{ktensor})  can be expanded in
perturbations around the background $\phi(t)=t$.
In synchronous gauge one finds
\begin{align}
  (T_{00})_P &= \rho_Q + \dot{\rho}_Q \pi + (\rho_Q + p_Q + 4M^4) \dot{\pi} \, ,\label{kT00} \\
  (T_{0i})_P &= (\rho_Q + p_Q) \partial_i\pi \, ,\label{kT0i} \\
  (T_{ij})_P &= a^2 \delta_{ij} \left[ p_Q + \dot{p}_Q \pi + (\rho_Q + p_Q)\dot{\pi} \right] + a^2 p_Q h_{ij} \label{kTij} \, .
\end{align}

One can do the same for the operators proportional to $\Mb^2$ and
$\hat{M}^3$ and show that they can also be written in the
perfect fluid form \eqref{perfectfluid}, by noting that
$u^\mu$ defined in eq.~(\ref{velocity}) can be written in several different ways,
\begin{equation}
u_\mu =
\frac{\partial_\mu X}{\sqrt{-(\partial X)^2}}
= \frac{\partial_\mu\square\phi}{\sqrt{-(\partial\square\phi)^2}}= \frac{\partial_\mu(\square\phi
+ 3H(\phi))}{\sqrt{-[\partial(\square\phi + 3H)]^2}} \, .
\end{equation}
Let us first compute the
contribution to the stress-energy tensor of the operator
proportional to $\Mb^2$, which reads\footnote{Here we take for simplicity $\Mb$ and $\hat{M}$ to
be constant.}
\begin{equation}
(T_{\mu\nu})_{\Mb^2} = \Mb^2\bigg[ -
2\partial_{(\mu}(\square\phi + 3H)\partial_{\nu)}\phi + 
\frac{1}{2}(\square\phi -
3H)(\square\phi + 3H)g_{\mu\nu}
+ g^{\alpha\beta}\partial_\alpha(\square\phi + 3H)\partial_\beta\phi g_{\mu\nu} \bigg]\,,
\end{equation}
where $_{(\mu \nu)}$ denotes symmetrization in $\mu$ and $\nu$.
The perfect fluid form  (\ref{perfectfluid})  can be obtained by recognizing
its contribution to the energy density and pressure, defined as
\begin{align}
\rho_{\Mb^2} &\equiv  \Mb^2\left[ g^{\mu\nu}\partial_\mu(\square\phi + 3H)\partial_\nu\phi
-\frac{1}{2}(\square\phi - 3H)(\square\phi
+ 3H) \right]
\, ,
\\
p_{\Mb^2} &\equiv  \Mb^2\left[g^{\mu\nu}\partial_\mu(\square\phi + 3H)\partial_\nu\phi + \frac{1}{2}(\square\phi - 3H)(\square\phi
+ 3H) \right]
\, .
\end{align}
In order to expand in perturbations let us first compute the
Laplace-Beltrami operator,
\begin{equation}
  \square \phi = \frac{1}{\sqrt{-g}} \de_\mu \left( \sqrt{-g} g^{\mu
      \nu} \de_\nu \phi \right)
  = -3H - \ddot{\pi} - 3H\dot{\pi} +
  \frac{\nabla^2 \pi}{a^2} - \frac{1}{2}\dot{h} \, .
\end{equation}
Then the stress-energy tensor becomes
\begin{equation}
\begin{split}
  (T_{\mu\nu})_{\Mb^2} = \bar{M}^2\bigg[&(\partial_t + 3H) \bigg(
  \ddot{\pi} + 3H\dot{\pi} -3\dot{H}\pi - \frac{\nabla^2 \pi}{a^2} +
  \frac{1}{2}\dot{h}\bigg)g_{\mu\nu}\\ &+ 2 \del_{0
    (\nu} \partial_{\mu)} \bigg(\ddot{\pi} + 3H\dot{\pi} -3\dot{H}\pi
  - \frac{\nabla^2 \pi}{a^2} +\frac{1}{2}\dot{h}\bigg) \bigg] \, .
\label{stressenergyMb}
\end{split}
\end{equation}

We now perform the same procedure for the operator proportional to $\hat{M}^3$. Its stress-energy tensor is
\begin{equation}
(T_{\mu\nu})_{\hat{M}^3} = \hat{M}^3\left[-(\square\phi +
3H)\partial_\mu\phi\partial_\nu\phi -\partial_{(\mu}X\partial_{\nu)}\phi - \frac{3H}{2}g_{\mu\nu}(X - 1)
+
\frac{1}{2}g_{\mu\nu}\partial^\alpha X\partial_\alpha\phi\right] \, ,
\end{equation}
and once again it can be written in the perfect fluid form by defining
\begin{align}
\rho_{\hat{M}^3} &\equiv \hat{M}^3\left[
\frac{1}{2} g^{\mu\nu} \partial_\mu X \partial_\nu\phi
-(\square\phi + 3H)X + \frac{3}{2}H(X - 1) \right] \, , \\
p_{\hat{M}^3} &\equiv \hat{M}^3\left[
 \frac{1}{2}g^{\mu\nu} \partial_\mu X \partial_\nu \phi
- \frac{3}{2}H(X - 1) \right] \, .
\end{align}
When expanded to first order in the perturbations, the stress-energy
tensor for the operator proportional to $\hat{M}^3$ takes the form
\begin{align}
  (T_{00})_{\hat{M}^3} &= \hat{M}^3 \left(6 H \dot{\pi} -
    3\dot{H}\pi - \frac{\nabla^2 \pi}{a^2} + \frac{1}{2}\dot{h}
  \right) \, , \label{T00Mh}\\
  (T_{0i})_{\hat{M}^3} &= - \hat{M}^3\de_i\dot{\pi} \, ,\\
  (T_{ij})_{\hat{M}^3} &= -\hat{M}^3a^2\delta_{ij}(\partial_t +
  3H)\dot{\pi}\, . \label{TijMh}
\end{align}

Using the notation of \cite{Ma:1995ey},
the stress-energy tensor of a perfect fluid in synchronous gauge can be written as
\begin{align}
  T_{00} &= \rho_Q + \del \rho_Q \, , \\
  T_{0i} &=  - a\, (\rho_Q + p_Q)  \partial_i \nabla^{-2} \theta_Q \, , \\
  T_{ij} &= (p_Q + \delta p_Q) a^2 \delta_{ij} + a^2 p_Q h_{ij}\, .
\end{align}
Notice that we defined $\theta_Q$ in the main text, below
eq.~(\ref{euler}), using conformal time \cite{Ma:1995ey}. Here
we are in cosmic time and this introduces the factor of $a$ in the equation for $T_{0i}$ above.
Thus, from equations \eqref{kT00}--\eqref{kTij},
\eqref{stressenergyMb}, and \eqref{T00Mh}--\eqref{TijMh} one gets,
for our complete stress-energy tensor,
\begin{align}
  \delta\rho_Q =& \dot{\rho}_Q\pi + (\rho_Q + p_Q + 4M^4)\dot{\pi} \nonumber \\
  & + \bar{M}^2\bigg[(\partial_t - 3H) \bigg( \ddot{\pi} + 3H\dot{\pi}
  -3\dot{H}\pi -
  \frac{\nabla^2 \pi}{a^2} + \frac{1}{2}\dot{h}\bigg)\bigg] \nonumber \\
  & + \hat{M}^3 \bigg( 6 H \dot{\pi} - 3 \dot{H} \pi - \frac{\nabla^2
    \pi}{a^2} + \frac{1}{2}\dot{h} \bigg) \, , \\
  a \, \theta_Q =& -\nabla^2 \pi +\frac{\bar{M}^2}{\rho_Q + p_Q}
  \nabla^2\bigg(-\ddot{\pi} - 3H\dot{\pi} + 3\dot{H}\pi +
  \frac{\nabla^2 \pi}{a^2} - \frac{1}{2}\dot{h}\bigg) +
  \frac{\hat{M}^3}{\rho_Q + p_Q}\nabla^2\dot{\pi} \, ,\\
  \delta p_Q =& \dot{p}_Q\pi + (\rho_Q + p_Q)\dot{\pi} \nonumber \\
  & + \bar{M^2}(\partial_t + 3H) \bigg(\ddot{\pi} + 3H\dot{\pi}
  -3\dot{H}\pi - \frac{\nabla^2 \pi}{a^2} + \frac{1}{2}\dot{h} \bigg)
  - \hat{M}^3(\partial_t + 3H)\dot{\pi} \, .
\end{align}


\section{General action in unitary gauge}
\label{app:unitary}

We wish to write the general action for the perturbations of a single
quintessence field minimally coupled to gravity with no direct
couplings to other fields.  Following ref.~\cite{StartingUniverse,EFTinflation}, we
choose a gauge where the scalar field perturbation is set to zero but
it appears as a scalar metric degree of freedom. In this gauge the
constant time hypersurfaces are equivalent to the uniform field
hypersurfaces.  In the fluid language, this implies that the velocity
is orthogonal to the constant time surfaces, $T^0_i=0$, that is why this gauge is
called `velocity orthogonal' \cite{Kodama:1985bj}. Using a particle physics terminology we
can also call it `unitary gauge', as all the degrees of
freedom are in the metric.
Notice that one can always parameterize the perturbations of the
scalar field as
\begin{equation}
\phi(t,\vec x) = \phi_0(t+\pi(t,\vec x))\, , \label{pi_def}
\end{equation}
where $\pi$ is the difference between the constant time and uniform field
hypersurfaces, so that the unitary gauge corresponds to $\pi=0$.

In order to find the effective action for quintessence, gravity and
other matter components described by the Lagrangian ${\cal L}_m$, we
write down all the terms that preserve the symmetries of the
system. Our choice of gauge breaks time diffeomorphism invariance
while preserving invariance under spatial diffeomorphisms. Thus, we
include linear and quadratic combinations of generic functions of time
$t$, the time-time component of the inverse metric $g^{00}$ and the
extrinsic curvature of the constant time hypersurfaces.  The
effective action up to second order in perturbations is
\cite{StartingUniverse,EFTinflation}:
\begin{equation}
\begin{split}
\label{unitaryaction}
  S = \int \ud^3 x \,\ud t \,\sqrt{-g} \Bigg[& \frac{\Mpl^2}{2} R +
  {\cal L}_m + c(t) g^{00} - \Lam(t)
  + \frac{M^4(t)}{2} (g^{00}+1)^2 - \frac{\Mb^2(t)}{2} \del K^2 \\
  &- \frac{\tilde{M}^2(t)}{2} \del K^i_{\;j} \del K^j_{\;i} - \frac{\hat{M}^3(t)}{2}
\del K (g^{00}+1)\Bigg] \, ,
\end{split}
\end{equation}
where $R$ is the Ricci scalar and $K_{ij}$ is the extrinsic curvature
of constant $t$ hypersurfaces, which at linear order reads
\begin{equation}
K_{ij} =  \frac{1}{2} \sqrt{- g^{00}} \left( \de_0 g_{ij} - \de_i g_{0 j} - \de_j g_{i 0} \right) \, ,
\end{equation}
and we have defined $\delta K_{ij} \equiv K_{ij} - a^2H\delta_{ij}$, and $\delta K
\equiv K^i_{\; i} - 3H$.

There are other operators that are invariant under spatial
diffeomorphisms that one would na\"{i}vely include in this action.
However, these operators are irrelevant at energy scales below the
cutoff $M \sim \bar M \sim \hat M$.  For instance, one could include
the operator $(\dot{g}^{00})^2$. This would give the term
$\ddot{\pi}^2$ in the final action, which indeed also appears as part
of the operator \eqref{Boxphi}, once expanded in the perturbations,
eq.~\eqref{barM}.  However, as explained in section~\ref{sec:hd}, for
frequencies smaller than the cutoff $\ddot \pi^2$ is negligible with
respect to $(\nabla^2 \pi)^2$ so that it can be ignored in the action.
For simplicity we will also ignore the operator proportional to
$\tilde{M}^2$ as it leads to terms qualitatively similar to those
proportional to $\Mb^2$.

One can easily fix the coefficients $c(t)$ and
$\Lambda(t)$ by computing the background stress-energy tensor. This gives
\begin{align}
  \rho_Q &= \Lam(t) - c(t) \, , \\
  p_Q &= - c(t) - \Lam(t) \, .
\end{align}
Finally, using these relations, we can write the
action~\eqref{unitaryaction} in unitary gauge in terms of the
background quantities $\rho_Q(t)$ and $p_Q(t)$,
\begin{equation}
\begin{split}
\label{unitaryaction2}
  S = \int \ud^4 x \,\sqrt{-g} \Bigg[& \frac{\Mpl^2}{2} R + {\cal L}_m
  + p_Q - \frac{1}{2} (\rho_Q + p_Q) (g^{00} + 1)
  + \frac{M^4(t)}{2} (g^{00}+1)^2 \\ &- \frac{\Mb^2(t)}{2} \del K^2
  -\frac{\hat{M}(t)^3}{2}\delta K(g^{00} + 1)\Bigg] \, .
\end{split}
\end{equation}

Now we want to rewrite this action in a gauge-invariant form. This can
be done by  performing the following time-coordinate transformation,
\begin{equation}
  t \to \tilde{t} = t + \pi(x) \, \qquad x^i \to \tilde{x}^i = x^i \, ,
\end{equation}
that reintroduces $\pi$ defined in eq.~(\ref{pi_def}).
The action for $\pi$ reads, up to second order,
\begin{equation}
\begin{split}
\label{actionpi}
  S = &\int \ud^4 x \,\sqrt{-g} \Bigg\{
p_Q + \dot{p}_Q \pi + \frac{1}{2} \ddot{p}_Q \pi^2 \\
  &- \frac{1}{2} (\rho_Q + p_Q) \left[ (g^{00} +1) - 2 \dot{\pi}
  + 2 (g^{00} + 1) \dot{\pi} - \dot{\pi}^2 + 2 g^{0i} \de_i \pi + \frac{(\nab \pi)^2}{a^2} \right]  \\
&- \frac{1}{2} (\dot{\rho}_Q + \dot{p}_Q) \pi \left[ (g^{00} +1) - 2 \dot{\pi} \right]
  + \frac{M^4(t)}{2} \left[ (g^{00} +1) - 2 \dot{\pi} \right] ^2 \\
  &- \frac{\Mb^2(t)}{2} \left( \del K - 3 \dot{H}\pi - \frac{\nab^2 \pi}{a^2}
  \right)^2 - \frac{\hat{M}(t)^3}{2} \left(\delta K - 3 \dot{H}\pi - \frac{\nabla^2
      \pi}{a^2}\right)\left[(g^{00}+1) - 2 \dot{\pi} \right]\Bigg\} \, ,
\end{split}
\end{equation}
while the part of the action containing
$R$ and ${\cal L}_m$ is
invariant under general diffeomorphisms.
We now choose to work in the synchronous gauge,
which is defined by the metric
\begin{equation}
 \ud s^2 = - \ud t^2 + a^2(t) (\del_{ij} + h_{ij}) \ud x^i \ud x^j \, .
\end{equation}
Using the notation of \cite{Ma:1995ey}, the two scalar degrees of freedom of $h_{ij}$ are its trace $h \equiv
\del^{ij} h_{ij}$ and $\eta$ which is defined by $\nabla^2h_{ij} \equiv \de_i\de_jh
+ 6( \de_i \de_j -
\frac{1}{3} \del_{ij}\nabla^2) \eta$.
Using this metric,
after integrating by parts and using the background continuity
equation $\dot{\rho}_Q + 3H(\rho_Q + p_Q) = 0$, the action (\ref{actionpi})
takes the form
\begin{equation}
\begin{split}
  S
 =&\int \ud^4x\,a^3\bigg\{ p_Q
  + \frac{1}{2}(\rho_Q + p_Q)\left[\dot{\pi}^2 - \frac{(\nab \pi)^2}{a^2}\right]
  + 2 M^4\dot{\pi}^2
   + \frac{3}{2}\dot{H}(\rho_Q + p_Q)\pi^2 - \frac{1}{2}(\rho_Q + p_Q)\dot{h}\pi \\
&  -\frac{\Mb^2}{2}\left( \frac{1}{2} \dot{h}
    -3 \dot{H}\pi - \frac{\nab^2\pi}{a^2} \right)^2
  +\hat{M}^3\dot{\pi}\left(\frac{1}{2}\dot{h} -3\dot{H}\pi - \frac{\nabla^2
      \pi}{a^2} \right)\bigg\} \, . \label{action}
\end{split}
\end{equation}

We can now compute the stress-energy tensor of quintessence
using the action \eqref{actionpi}. Expanding in the perturbations, its
components read
\begin{align}
  T_{00} &= \rho_Q + \left(\rho_Q + p_Q + 4 M^4 \right) \dot{\pi} +
  \dot{\rho}_Q \pi \nonumber \\
  &\phantom{=} -3H\Mb^2\bigg(\frac{1}{2}\dot{h} - 3 \dot{H}\pi -
  \frac{\nabla^2}{a^2}\pi\bigg) + \hat{M}^3\bigg(\frac{1}{2}\dot{h} -
  3 \dot{H}\pi -
  \frac{\nabla^2}{a^2}\pi + 3H\dot{\pi} \bigg)  \, ,\\
  T_{0i} &= (\rho_Q + p_Q) \de_i \pi + \Mb^2 \partial_i\bigg(
  \frac{1}{2} \dot{h} -3 \dot{H}\pi -\frac{\nab^2 \pi}{a^2} \bigg) -
  \hat{M}^3\partial_i\dot{\pi} \, ,\\
  T_{ij} &= p_Q a^2 \del_{ij} + \left[ \dot{p}_Q \pi + \left( \rho_Q +
      p_Q \right) \dot \pi \right] a^2 \del_{ij} + p_Q
  a^2 h_{ij} \nonumber \\
  &\phantom{=} + 2\Mb^2 a^2 \delta_{ij} (\de_0 + 3 H) \bigg( \frac{1}{2}\dot{h} -
  3 \dot{H}\pi - \frac{\nab^2 \pi}{a^2} \bigg) -
  2\hat{M}^3a^2\del_{ij}(\partial_0 + 3H)\dot\pi \, .
\end{align}

In the main body of this paper we have constructed the action for
$\pi$ using a different procedure from the one presented here.
Indeed, we started from the action of $k$-essence,
eq.~(\ref{kessence}), and we added the two $\phi$-dependent higher
derivative operators $-\bar M^2(\phi)[\square\phi + 3H(\phi)]^2/2$
and $- \hat M^3(\phi) [\square\phi + 3H(\phi)](X - 1)/2$, that do not
change the background equations of motion.
Also the action (\ref{action}) can be constructed similarly.
First of all, note that the first line of
eq.~(\ref{action}) is the action for $k$-essence. Indeed, it is
equivalent to the action (\ref{kessencefull}),
which was found by expanding the $k$-essence action (\ref{kaction}) in
terms of $\pi$. The second line of eq.~(\ref{action}) can be constructed by
noting that the extrinsic
curvature of the hypersurfaces of constant $\phi$, defined as
\begin{equation}
K^\mu_{\phantom{\mu}\nu} \equiv -(g^\rho_{\phantom{\rho}\nu} + u^\rho
u_\nu) \nabla_\rho u^\mu \; ,
\end{equation}
where $u^\mu$ is the unit vector orthogonal to $\phi$
defined in (\ref{velocity}), can be
rewritten as
\begin{equation}
\label{Kphi}
K^\mu_{\phantom{\mu}\nu} =
\frac{1}{\sqrt{-(\partial\phi)^2}}\Bigg[\nabla^\mu\partial_\nu\phi +
\frac{\partial^\mu\phi\partial^\rho\phi}{-(\partial\phi)^2}\nabla_\rho\partial_\nu\phi\Bigg]
\,.
\end{equation}
When expanded around the background solution $\phi=t$, then $\delta K
\equiv K_{\
  \mu}^\mu - 3H(\phi)$ reads\footnote{One may wonder why terms of the form
$\dot{\pi}\nabla^2\pi$ do not appear in $\delta K^2$, while they do
appear in $(\square\phi + 3H)^2$. As seen in eq.~\eqref{Kphi},
a time diffeomorphism does not change the extrinsic
curvature of constant $\phi$ hypersurfaces. Thus, $\delta K$ does not contain $\dot \pi$ and
$\dot \pi \nabla^2\pi$ is not generated by $\delta K^2$.}
\begin{equation}
\delta K = -3\dot{H}\pi - \frac{\nabla^2}{a^2}\pi +
\frac{1}{2}\dot{h} \, .
\end{equation}
Thus the action \eqref{action} can be constructed by simply adding to the
$k$-essence action the two operators  $-\bar M^2(\phi) \delta
K^2/2$ and $-\hat M^3 (\phi) \delta K(g^{00} + 1)/2$, that do not
change the background solution.


\section{Modification of gravity with $\hat M$}
\label{app:modgrav}

In the main text we studied the effects of the operator $\hat M$,
focusing on stability and on the phenomenology at cosmological
scales. In analogy to what happens for the Ghost Condensate, we expect
that this operator will also be relevant at short scales, inducing a
modification of gravity. In this appendix we perform a preliminary
analysis, restricted to linear perturbations only, although
the non-linear dynamics has been shown to be relevant and quite rich in the Ghost Condensate case (see for example \cite{ArkaniHamed:2005gu}).  To simplify
the analysis we set $\rho_Q + p_Q=0$ and $\bar M=0$. Although the
background quintessence stress-energy tensor is the one of the
cosmological constant, there is still a propagating scalar degree of
freedom. Its mixing with gravity induces a deviation from General
Relativity; indeed the Ghost Condensate was originally proposed as a
consistent modification of gravity in the infrared. The simplest
setting to study this modification of gravity is in the Newtonian
regime $\omega^2 \ll k^2$ around Minkowski spacetime, where the new
scalar degree of freedom modifies the Newtonian potential $\Phi$. For
this purpose we will closely follow the discussion done in
\cite{GhostCond} for the Ghost Condensate case, i.e., for the operator
$\bar M$.

Working in Newtonian gauge with $\Psi = \Phi$, the metric is $\ud s^2 = - (1+ 2 \Phi)
\ud t^2 + (1- 2 \Phi) \ud \vec x^2$ and the quadratic Lagrangian for $\pi$ and
$\Phi$ reads
\begin{equation}
{\cal L}= - \Mp^2 (\nabla\Phi)^2 + 2 M^4 (\Phi - \dot\pi)^2 + \hat M^3 (\Phi-
\dot\pi) (4 \dot\Phi - \ddot\pi+ \nabla^2\pi) \;.
\end{equation}
Let us first assume that $\hat M$ is time independent. Dropping total
derivatives and terms which are negligible in the limit $\omega^2 \ll
k^2$, we are left with
\begin{equation}
{\cal L}= - \Mp^2 (\nabla\Phi)^2 + 2 M^4 (\Phi - \dot\pi)^2 + \hat M^3 \pi \nabla^2\Phi \;.
\end{equation}

In terms of the canonically normalized fields $\pi_c \equiv 2 M^2 \pi$
and $\Phi_c \equiv \sqrt{2} \Mp \Phi$, the Lagrangian in Fourier space
can be written as
\begin{equation} {\cal L} =
\frac{1}{2}
\begin{pmatrix} \pi_c \! \! &  \! \! \Phi_c
\end{pmatrix}
{\cal M}
\begin{pmatrix} \pi_c \cr \Phi_c
\end{pmatrix}\;,
\end{equation}
with
\begin{equation}
{\cal M} \equiv
\begin{pmatrix}
\omega^2 & -i  \omega {\sqrt{2}M^2} /{\Mp}   - k^2 {\hat M^3 }/({ 2 \sqrt{2} M^2 \Mp})  \cr i  \omega {\sqrt{2}M^2} /{\Mp}  -  k^2 {\hat M^3 }/({2 \sqrt{2} M^2 \Mp})   & -k^2 + {2M^4}/{\Mp^2}
\end{pmatrix}
\label{matrix} \;.
\end{equation}
Setting to zero the determinant of this matrix gives the dispersion
relation
\begin{equation}
\omega^2 = - \frac{\hat M^6}{8 M^4 \Mp^2} k^2\;,
\end{equation}
which reproduces the Jeans instability already shown in eq.~(\ref{Mhat_pi2}).
The Jeans instability arises from the non-diagonal (mixing) term and
it is thus proportional to $\hat M^6$ instead of $\hat M^3$.

To study the corrections to the Newtonian theory, one can look at the
propagator of $\Phi$ that is the $\langle\Phi,\Phi\rangle$ entry of ${\cal{M}}^{-1}$. This can be written as
\begin{equation}
-\frac{1}{k^2} \cdot \left[ 1 - \frac{k^2 \hat
      M^6}{ 8M^4 \Mp^2}  \cdot \frac1{\omega^2 + k^2 \hat M^6/(8 M^4 \Mp^2)} \right]\;,
\end{equation}
where the term $-1/k^2$ is simply the standard Newtonian
propagator. As expected, a
substantial deviation requires, for a given distance $k^{-1}$, a
sufficient time $\omega^{-1}$ for the Jeans instability to develop, i.e.,
\begin{equation}
\omega^2 \lesssim \frac{\hat M^6}{8 M^4 \Mp^2} k^2 \;.
\end{equation}

We can now consider the case of a time dependent $\hat M$. In this way
we introduce new terms in the action that were previously dropped because
total derivatives. The same happens if we had considered a time dependent
spatial metric, but here we stick to Minkowski for simplicity. The
matrix ${\cal M}$ becomes
\begin{equation}
{\cal M} \equiv
\begin{pmatrix}
\omega^2 - k^2 H \hat M^3/4 M^4 & -i  \omega {\sqrt{2}M^2} /{\Mp}   - k^2 {\hat M^3 }/({2
  \sqrt{2} M^2 \Mp})  \cr i  \omega {\sqrt{2}M^2} /{\Mp}  -  k^2 {\hat M^3 }/({2
  \sqrt{2} M^2 \Mp})   & -k^2 + {2M^4}/{\Mp^2} - 2 H \hat M^3/\Mp^2
\end{pmatrix}
\label{matrix2} \;,
\end{equation}
where $H$ is the typical rate of variation of $\hat M^3$, $\dot {\hat
M}^3 = H \hat M^3$ (if the time dependence is induced by the metric
this becomes the Hubble rate). Computing the determinant and
restricting to frequencies much larger than $H$ we get the dispersion
relation
\begin{equation}
\omega^2 = \frac{H \hat M^3}{4 M^4} k^2 - \frac{\hat M^6}{8 M^4 \Mp^2} k^2\;,
\end{equation}
which correctly matches eq.~(\ref{Mhat_pi2}). The value of $\hat M$
can be chosen to avoid the Jeans instability and have a healthy
dispersion relation. The propagator
becomes
\begin{equation}
-\frac{1}{k^2} \cdot \left[ 1 - \frac{k^2 \hat
      M^6}{ 8M^4 \Mp^2}  \cdot \frac1{\omega^2 -k^2 \hat M^3 H/(4 M^4)+ k^2 \hat M^6/(8 M^4 \Mp^2)} \right]\;.
\end{equation}

The scalar degree of freedom induces a $1/r$ force which adds to the
Newton law: this force, however, propagates at a very small
speed
\begin{equation}
  c_s^2 \approx \frac{\hat M^3 H}{4M^4}.
\end{equation}
Given the absence of a Jeans instability, the modification of gravity
induced by $\hat M$ is very different at linear and non-linear level
with respect to the Ghost Condensate case
\cite{GhostCond,ArkaniHamed:2005gu}. More work is needed to understand
the constraints on $\hat M$ coming from the modifications of gravity that
it produces.


\footnotesize
\parskip 0pt

\begin{thebibliography}{99}


\bibitem{StartingUniverse}
  P.~Creminelli, M.~A.~Luty, A.~Nicolis and L.~Senatore,
  ``Starting the universe: Stable violation of the null energy condition and non-standard cosmologies,''
  JHEP {\bf 0612} (2006) 080
  [arXiv:hep-th/0606090].

\bibitem{EFTinflation}
  C.~Cheung, P.~Creminelli, A.~L.~Fitzpatrick, J.~Kaplan and L.~Senatore,
  ``The Effective Field Theory of Inflation,''
  JHEP {\bf 0803} (2008) 014
  [arXiv:0709.0293 [hep-th]].

\bibitem{Albrecht:2006um}
  A.~Albrecht {\it et al.},
  ``Report of the Dark Energy Task Force,''
  arXiv:astro-ph/0609591.

\bibitem{GhostCond}
  N.~Arkani-Hamed, H.~C.~Cheng, M.~A.~Luty and S.~Mukohyama,
  ``Ghost condensation and a consistent infrared modification of gravity,''
  JHEP {\bf 0405} (2004) 074
  [arXiv:hep-th/0312099].

\bibitem{Li:2005fm}
  M.~z.~Li, B.~Feng and X.~m.~Zhang,
  ``A single scalar field model of dark energy with equation of state  crossing
  -1,''
  JCAP {\bf 0512}, 002 (2005)
  [arXiv:hep-ph/0503268].

\bibitem{Feng:2004ad}
  B.~Feng, X.~L.~Wang and X.~M.~Zhang,
  ``Dark Energy Constraints from the Cosmic Age and Supernova,''
  Phys.\ Lett.\  B {\bf 607}, 35 (2005)
  [arXiv:astro-ph/0404224].

\bibitem{HuPhantom}
  W.~Hu,
  ``Crossing the phantom divide: Dark energy internal degrees of freedom,''
  Phys.\ Rev.\  D {\bf 71} (2005) 047301
  [arXiv:astro-ph/0410680].

\bibitem{Vikman:2004dc}
  A.~Vikman,
  ``Can dark energy evolve to the phantom?,''
  Phys.\ Rev.\  D {\bf 71}, 023515 (2005)
  [arXiv:astro-ph/0407107].

\bibitem{CaldwellDoran}
  R.~R.~Caldwell and M.~Doran,
  ``Dark-energy evolution across the cosmological-constant boundary,''
  Phys.\ Rev.\  D {\bf 72} (2005) 043527
  [arXiv:astro-ph/0501104].

\bibitem{ArmendarizPicon:1999rj}
  C.~Armendariz-Picon, T.~Damour and V.~F.~Mukhanov,
  ``k-inflation,''
  Phys.\ Lett.\  B {\bf 458}, 209 (1999)
  [arXiv:hep-th/9904075].


\bibitem{ArmendarizPicon:2000dh}
  C.~Armendariz-Picon, V.~F.~Mukhanov and P.~J.~Steinhardt,
  ``A dynamical solution to the problem of a small cosmological constant  and
  late-time cosmic acceleration,''
  Phys.\ Rev.\ Lett.\  {\bf 85}, 4438 (2000)
  [arXiv:astro-ph/0004134].

\bibitem{Cline:2003gs}
  J.~M.~Cline, S.~Jeon and G.~D.~Moore,
  ``The phantom menaced: Constraints on low-energy effective ghosts,''
  Phys.\ Rev.\  D {\bf 70}, 043543 (2004)
  [arXiv:hep-ph/0311312].

\bibitem{Holdom:2004yx}
  B.~Holdom,
  ``Accelerated expansion and the Goldstone ghost,''
  JHEP {\bf 0407}, 063 (2004)
  [arXiv:hep-th/0404109].

\bibitem{Caldwell:1999ew}
  R.~R.~Caldwell,
  ``A Phantom Menace?,''
  Phys.\ Lett.\  B {\bf 545}, 23 (2002)
  [arXiv:astro-ph/9908168].

\bibitem{Garriga:1999vw}
  J.~Garriga and V.~F.~Mukhanov,
  ``Perturbations in k-inflation,''
  Phys.\ Lett.\  B {\bf 458}, 219 (1999)
  [arXiv:hep-th/9904176].


\bibitem{Bonvin:2007mw}
  C.~Bonvin, C.~Caprini and R.~Durrer,
  ``Superluminal motion and closed signal curves,''
  arXiv:0706.1538 [astro-ph].

\bibitem{Babichev:2007dw}
  E.~Babichev, V.~Mukhanov and A.~Vikman,
  ``k-Essence, superluminal propagation, causality and emergent geometry,''
  JHEP {\bf 0802}, 101 (2008)
  [arXiv:0708.0561 [hep-th]].

\bibitem{Adams:2006sv}
  A.~Adams, N.~Arkani-Hamed, S.~Dubovsky, A.~Nicolis and R.~Rattazzi,
  ``Causality, analyticity and an IR obstruction to UV completion,''
  JHEP {\bf 0610}, 014 (2006)
  [arXiv:hep-th/0602178].

\bibitem{Hsu:2004vr}
  S.~D.~H.~Hsu, A.~Jenkins and M.~B.~Wise,
  ``Gradient instability for $w<-1$,''
  Phys.\ Lett.\  B {\bf 597}, 270 (2004)
  [arXiv:astro-ph/0406043].

\bibitem{Senatore:2004rj}
  L.~Senatore,
  ``Tilted ghost inflation,''
  Phys.\ Rev.\  D {\bf 71}, 043512 (2005)
  [arXiv:astro-ph/0406187].

\bibitem{Ma:1995ey}
  C.~P.~Ma and E.~Bertschinger,
  ``Cosmological perturbation theory in the synchronous and conformal Newtonian
  gauges,''
  Astrophys.\ J.\  {\bf 455}, 7 (1995)
  [arXiv:astro-ph/9506072].

\bibitem{ArkaniHamed:2005gu}
  N.~Arkani-Hamed, H.~C.~Cheng, M.~A.~Luty, S.~Mukohyama and T.~Wiseman,
  ``Dynamics of gravity in a Higgs phase,''
  JHEP {\bf 0701}, 036 (2007)
  [arXiv:hep-ph/0507120].

\bibitem{Mukohyama:2006be}
  S.~Mukohyama,
  ``Accelerating universe and cosmological perturbation in the ghost
  condensate,''
  JCAP {\bf 0610}, 011 (2006)
  [arXiv:hep-th/0607181].

\bibitem{Dave:2002mn}
  R.~Dave, R.~R.~Caldwell and P.~J.~Steinhardt,
  ``Sensitivity of the cosmic microwave background anisotropy to initial
  conditions in quintessence cosmology,''
  Phys.\ Rev.\  D {\bf 66}, 023516 (2002)
  [arXiv:astro-ph/0206372].

\bibitem{Erickson:2001bq}
  J.~K.~Erickson, R.~R.~Caldwell, P.~J.~Steinhardt, C.~Armendariz-Picon and V.~F.~Mukhanov,
  ``Measuring the speed of sound of quintessence,''
  Phys.\ Rev.\ Lett.\  {\bf 88}, 121301 (2002)
  [arXiv:astro-ph/0112438].

\bibitem{DeDeo:2003te}
  S.~DeDeo, R.~R.~Caldwell and P.~J.~Steinhardt,
  ``Effects of the sound speed of quintessence on the microwave background  and
  large scale structure,''
  Phys.\ Rev.\  D {\bf 67}, 103509 (2003)
  [Erratum-ibid.\  D {\bf 69}, 129902 (2004)]
  [arXiv:astro-ph/0301284].

\bibitem{Weller:2003hw}
  J.~Weller and A.~M.~Lewis,
  ``Large Scale Cosmic Microwave Background Anisotropies and Dark Energy,''
  Mon.\ Not.\ Roy.\ Astron.\ Soc.\  {\bf 346}, 987 (2003)
  [arXiv:astro-ph/0307104].

\bibitem{BeanDore}
  R.~Bean and O.~Dore,
  ``Probing dark energy perturbations: the dark energy equation of state and
  speed of sound as measured by WMAP,''
  Phys.\ Rev.\  D {\bf 69}, 083503 (2004)
  [arXiv:astro-ph/0307100].

\bibitem{Kodama:1985bj}
  H.~Kodama and M.~Sasaki,
  ``Cosmological Perturbation Theory,''
  Prog.\ Theor.\ Phys.\ Suppl.\  {\bf 78} (1984) 1.

\bibitem{Hannestad:2005ak}
  S.~Hannestad,
  ``Constraints on the sound speed of dark energy,''
  Phys.\ Rev.\  D {\bf 71}, 103519 (2005)
  [arXiv:astro-ph/0504017].

\bibitem{Takada:2006xs}
  M.~Takada,
  ``Can A Galaxy Redshift Survey Measure Dark Energy Clustering?,''
  Phys.\ Rev.\  D {\bf 74}, 043505 (2006)
  [arXiv:astro-ph/0606533].

\bibitem{TorresRodriguez:2007mk}
  A.~Torres-Rodriguez and C.~M.~Cress,
  ``Constraining the Nature of Dark Energy using the SKA,''
  Mon.\ Not.\ Roy.\ Astron.\ Soc.\  {\bf 376}, 1831 (2007)
  [arXiv:astro-ph/0702113].

\bibitem{Hu:2004yd}
  W.~Hu and R.~Scranton,
  ``Measuring Dark Energy Clustering with CMB-Galaxy Correlations,''
  Phys.\ Rev.\  D {\bf 70}, 123002 (2004)
  [arXiv:astro-ph/0408456].


\bibitem{Corasaniti:2005pq}
  P.~S.~Corasaniti, T.~Giannantonio and A.~Melchiorri,
  ``Constraining dark energy with cross-correlated CMB and Large Scale
  Structure data,''
  Phys.\ Rev.\  D {\bf 71}, 123521 (2005)
  [arXiv:astro-ph/0504115].

\bibitem{Polchinski:1992ed}
  J.~Polchinski,
  ``Effective Field Theory And The Fermi Surface,''
  arXiv:hep-th/9210046.

\bibitem{Simon:1990jn}
  J.~Z.~Simon,
  ``The Stability of flat space, semiclassical gravity, and higher
  derivatives,''
  Phys.\ Rev.\  D {\bf 43}, 3308 (1991).

\bibitem{Simon:1990ic}
  J.~Z.~Simon,
  ``Higher derivative Lagrangians, non-locality, problems and
  solutions,''
  Phys.\ Rev.\  D {\bf 41}, 3720 (1990).

\bibitem{Kallosh07}
  R.~Kallosh, J.~U.~Kang, A.~Linde and V.~Mukhanov,
  ``The New Ekpyrotic Ghost,''
  JCAP {\bf 0804} (2008) 018
  [arXiv:0712.2040 [hep-th]].

\bibitem{Weinberg08}
  S.~Weinberg,
  ``Effective Field Theory for Inflation,''
  Phys.\ Rev.\  D {\bf 77}, 123541 (2008)
  [arXiv:0804.4291 [hep-th]].


\bibitem{ArkaniHamed:2007ky}
  N.~Arkani-Hamed, S.~Dubovsky, A.~Nicolis, E.~Trincherini and G.~Villadoro,
  ``A Measure of de Sitter Entropy and Eternal Inflation,''
  JHEP {\bf 0705}, 055 (2007)
  [arXiv:0704.1814 [hep-th]].

\end{thebibliography}
\end{document}